\documentclass[aps,10pt,pre,showpacs,preprintnumbers,nofootinbib,twocolumn]{revtex4-1}%

\usepackage{amsmath,amssymb,amsfonts,amsthm}
\usepackage{bm,bbm}
\usepackage{graphicx}
\usepackage{esint}
\usepackage{color}
\usepackage[breaklinks]{hyperref}

\newcommand{\be}{\begin{equation}}
\newcommand{\ee}{\end{equation}}
\newcommand{\ba}{\begin{eqnarray}}
\newcommand{\ea}{\end{eqnarray}}
\def\bs{\begin{subequations}}
\def\es{\end{subequations}}
\def\a{\alpha}
\def\b{\beta}
\def\de{\delta}
\def\g{\gamma}
\def\la{\lambda}

\def\e{\epsilon}
\def\ve{\varepsilon}

\def\s{\sigma}
\def\vr{\varrho}
\def\vp{\varphi}
\def\N{\nabla}

\def\cD{{\cal D}}

\def\cI{{\cal I}}

\def\cK{{\cal K}}
\def\cL{{\cal L}}
\def\cM{{\cal M}}
\def\cN{{\cal N}}

\def\cP{{\cal P}}

\def\cS{{\cal S}}

\def\cV{{\cal V}}
\def\bE{\mathbbm{e}}

\def\ds{d_{\rm S}}
\def\dh{d_{\rm H}}
\def\dw{d_{\rm W}}

\def\p{\partial}
\def\bp{\bar{\partial}}
\def\B{\Box}
\newcommand{\Eq}[1]{(\ref{#1})}
\def\com{\color{magenta}}
\def\cob{\color{blue}}
\newcommand{\oarX}[1]{\href{http://arxiv.org/abs/#1}{{\ttfamily\com #1}}}
\newcommand{\arX}[1]{\href{http://arxiv.org/abs/#1}{{\ttfamily\com arXiv:#1}}}
\newcommand{\doin}[5]{\href{http://dx.doi.org/#1}{\cob #2 {\bf #3}, #4 (#5)}}
\newcommand{\doij}[5]{\href{http://dx.doi.org/#1}{\cob #2 #3 (#5) #4}}
\newcommand{\ndoin}[5]{\href{#1}{\cob #2 {\bf #3}, #4 (#5)}}
\newcommand{\tia}[1]{}

\def\rme{e}
\def\rmd{d}
\def\rmi{i}

\begin{document}

\title{Diffusion in multiscale spacetimes}

\author{Gianluca Calcagni}
\email{calcagni@iem.cfmac.csic.es; Present address: Instituto de Estructura de la Materia, CSIC, Serrano 121, 28006 Madrid, Spain.}
\affiliation{Max Planck Institute for Gravitational Physics (Albert Einstein Institute)\\
Am M\"uhlenberg 1, D-14476 Golm, Germany}

\date{May 21, 2012}

\begin{abstract}
We study diffusion processes in anomalous spacetimes regarded as models of quantum geometry. Several types of diffusion equation and their solutions are presented and the associated stochastic processes are identified. These results are partly based on the literature in probability and percolation theory but their physical interpretation here is different since they apply to quantum spacetime itself. The case of multiscale (in particular, multifractal) spacetimes is then considered through a number of examples and the most general spectral-dimension profile of multifractional spaces is constructed.
\end{abstract}



\preprint{\doin{10.1103/PhysRevE.87.012123}{Phys.\ Rev.\ E}{87}{012123}{2013} [\arX{1205.5046}]}
\preprint{AEI-2012-043}

\maketitle


\section{Introduction}

Although many different theories of quantum gravity have been proposed to accommodate the gravitational interaction and quantum mechanics in a unified framework, certain features seem to be universal and deeply connected with one of the most desired properties all these candidates should possess: namely, the absence or control of divergences at small scales. One of these features is dimensional reduction, also known as dimensional flow \cite{tH93,Car09,Car10}. In general, one of the indicators characterizing quantum geometry, the spectral dimension $\ds$ of spacetime, changes with the scale, running from $\ds\lesssim 2$ (or exactly $\ds=2$) in the ultraviolet (UV) to the usual, classical value $\ds\sim 4$ in the infrared (IR). Numerical and analytic examples can be found in causal dynamical triangulations (CDT) \cite{AJL4,BeH}, random combs \cite{DJW1,AGW} and random multigraphs \cite{GWZ1,GWZ2} (both sharing some properties with CDT), quantum Einstein gravity (QEG, also called asymptotic safety) \cite{LaR5,ReS11}, spin foams \cite{Mod08,CaM,MPM,COT}, Ho\v{r}ava--Lifshitz gravity \cite{Hor3,SVW1}, noncommutative geometry at the fundamental \cite{Con06,CCM} and effective levels \cite{Ben08,ACOS,AA}, field theory on multifractal spacetimes \cite{fra1,fra2,fra3} (in particular, in the realization within multifractional geometry \cite{fra4,frc1,frc2,frc3,fra6,frc4}), and nonlocal super-renormalizable quantum gravity \cite{Mod11,AMM,Mod12}.

The multifractional framework is rather effective in describing geometric and physical features of quantum gravity at large. Therefore, it can be regarded either as an independent proposal for a fundamental theory or an effective framework wherein to better understand the multiscale geometry of the other approaches (examples are \cite{ACOS,fra7,CES}). For this reason, we believe it is important to exploit the tools available in multifractional spacetimes as much as possible. Due to the young age of the proposal, these tools are largely unexplored and the purpose of this paper is to continue the investigation initiated in Refs.\ \cite{fra4,frc1,frc2,frc3}, in the meanwhile improving the understanding of dimensional flow as a general phenomenon of quantum geometry.

To begin with, we need to refine the discussion on diffusion processes of Ref.\ \cite{frc1}. Systems with anomalous dimension are described by fractional versions of the standard diffusion equation, while multiscale systems are obtained by further but simple extensions. In Ref.\ \cite{fra6}, which is a condensed exposition of the contents of Secs.\ \ref{f4} and \ref{f6}, the following was pointed out. Once the forms of the Laplacian and of the diffusion operator are determined independently (e.g., by phenomenology), the requirement that the solution of the diffusion equation be non-negative definite allows one to fix other details of the diffusion equation, such as the presence of source terms, and to give a probabilistic interpretation to the diffusion process. Associating diffusion with a stochastic process is an important step towards the understanding of the physical properties of the quantum geometry under scrutiny and of dimensional flow. Multiscale cases are more complicated than those with fixed dimensionality, but this does not prevent the construction of systems with a composite scale hierarchy. Examples exist in the literature of probability and chaos theory that can be applied to quantum-gravity scenarios in the present interpretation. We shall also construct the profile of the spectral dimension for a multifractional spacetime with a finite but arbitrary number of characteristic scales. The author had the single-scale case in mind when presenting the main features of fractal and fractional field theory in Refs.\ \cite{fra1,fra2,fra3,fra4,frc2}, but multiscale extensions also have direct applications to quantum gravity, like the two-scale profile for asymptotic safety \cite{fra6,fra7}.

The plan of the paper is as follows. Section \ref{unun} introduces the triple issue of universality, robustness, and uniqueness of diffusion and geometric properties in quantum gravity, and motivates why we expect and study nonstandard diffusion equations generated by quantum-geometric effects. The main geometric indicators of continuum models of fractal spacetimes are reviewed in Sec.\ \ref{cfsf} with some additional material. After the extension of fractional measures to bilateral worlds \cite{frc3}, the calculation of the spectral dimension of Ref.\ \cite{frc1}, with further comments on the diffusion probability, can be generalized to these cases and to fractional Laplacians. In Secs.\ \ref{f2}--\ref{f2b} we exploit the recent construction of a momentum transform in fractional spaces \cite{frc3} and include a class of fractional Laplacians not considered in Refs.\ \cite{frc1,frc2}. The classification of different types of diffusion is heavily based upon known results in the literature on probability theory, stochastic processes, and diffusion, although here we give the subject a different spin by identifying the diffusion medium with spacetime itself (Sec.\ \ref{f4}). Control of multifractional spacetimes is greatly improved in Sec.\ \ref{f6}, where dimensional flow is treated analytically. Section \ref{disc} is devoted to conclusions.


\section{Universality, robustness, and uniqueness}\label{unun}

To probe the local geometry of a given classical spacetime, there is a standard recipe founded upon a diffusion equation. One places a pointwise test particle on the Euclidean (i.e., imaginary-time) version of spacetime and lets it diffuse from point $x$ to point $x'$, parametrizing the diffusion with an external variable $\s$. It is common to assume the diffusion equation \cite{Fic55,Ein05,Smo06}
\be\label{dife}
\left(\p_\s-\N_x^2\right) P(x,x',\s)=0\,,\qquad P(x,x',0)=\delta(x-x')\,,
\ee
where the initial condition states the nonextension of the probe. The parameter $\s\geq 0$ acts as an abstract ``time'' variable via the diffusion operator $\p_\s$, an ordinary first-order derivative. The Laplacian $\N^2$ (acting on the variable $x$), called more generally the spatial generator in probability theory, is a second-order differential operator.

Quantum geometry can emerge either by definition of a nonstandard texture of spacetime (as in noncommutative and multifractional spaces and in Ho\v{r}ava--Lifshitz theory) or from the discretization or quantization of the gravitational interaction (as in CDT and asymptotic safety), or for both reasons. In all cases, the spectral dimension $\ds$ becomes anomalous, i.e., different from the topological dimension, and not because of curvature effects which exist already at the classical level. Since $\ds$ stems from the diffusion equation, quantum geometry effectively modifies either the diffusion operator $\p_\s$, or the Laplacian $\N^2_x$, or the initial condition $P(x,x',0)$, or the three of them. To understand the origin of such deformations of Eq.\ \Eq{dife}, we also quote some models where this happens.
\begin{itemize}
\item {\bf Modification of $\p_\s$.} Diffusion ``time'' acquires an anomalous scaling and the diffusion operator becomes fractional (see below), signaling the emergence of a memory effect along the diffusion flow. In the presence of one or more fundamental quantum scales, the new operator is in fact a sum of operators of different orders. To the best of our knowledge, there are only two concrete examples of a quantum-gravity diffusion equation with (multi)fractional diffusion operator: multifractional spacetimes \cite{frc1} (but as an optional construction) and, perhaps more interestingly for the {habitu\'es} of the field, maybe also quantum Einstein gravity \cite{CES}. There, the deformation of $\p_\s$ is realized because the cut-off scale of the theory is not identified with the physical momentum as usual but, as a powerful alternative, with diffusion time; the anomalous scaling is due to the renormalization group flow realizing asymptotic safety in the UV, while the presence of several scales (and, hence, of several diffusion operators in the same equation) is guaranteed by the type of action. This liberty in the cut-off identification is part of the question of universality and robustness outlined below. Also, a change in the diffusion operator can be made equivalent (from the point of view of the asymptotic scaling of the variance) to a change of spatial generator, thanks to a duality between diffusion equations which we will introduce later. Finally, the relation between number of operators, number of scales, and scale-dependence of the spectral dimension is not a subject much explored in quantum gravity (but see \cite{SVW2}) and will be extensively analyzed here.
\item {\bf Modification of $\N^2_x$.} As the spacetime texture gets modified (by quantum-gravity effects or by a nonstandard \emph{Ansatz} of background geometry), the differential structure also changes, and with that the notion of Laplacian. Noncommutative geometry and multifractional spacetimes are two instances: in both cases, calculus itself and the measure of spacetime undergo deep revisions from first principles. Qualitatively, asymptotic safety sports higher-order Laplacians (``dual'' to lower-order diffusion operators) because of the anomalous scaling of the metric within (which survives in inertial frames). In Ho\v{r}ava--Lifshitz gravity, higher-order Laplacians arise by a simple power-counting argument of the quantum theory \cite{Hor3}. In CDT, spin foams, and simplicial gravity in general, the deformation of the Laplacian is often not explicit but one can understand it as coming from the semiclassical continuum limit of a discretized geometry where multiple scales are introduced \cite{COT}. Dilaton black holes are another example of diffusion equation with modified Laplacian \cite{CaG}.
\item {\bf Modification of $P(x,x',0)$.} On a manifold with anomalous properties such as in a quantum setting, the notion of ``point particle'' can undergo a change to adapt to the different geometric background. In other words, the Dirac distribution $\delta(x-x')$ no longer plays the role of the ``delta'' and must be replaced by a different distribution. Two cases where this happens are quantum manifolds with a minimal length, where pointwise objects are smeared to Gaussians \cite{MoN,SSN}, and, again, multifractional spacetimes, where the Dirac distribution must be substituted by its fractional generalization \cite{frc1,frc3}.
\end{itemize}
A classification of the possible diffusion equations should help to understand the physics beyond these and other models of quantum gravity. To carve concrete inroads into multiscale quantum geometry, it is instructive to focus one's attention on the characteristics of dimensional flow, starting from the most general ones. All models of quantum gravity can be equipped with a diffusion equation and, hence, a sensible definition of the effective dimension of spacetime (the spectral dimension $\ds$). Then, one can  pose a series of questions grouped into three main issues:

\emph{Universality}: Are there geometric features common to all or the great majority of the approaches to quantum gravity? If not, are there features which are general to many models although not universal?

\emph{Robustness}: Are there features of dimensional flow insensitive to the fine details?

\emph{Uniqueness}: Given a model of quantum gravity, is the diffusion equation (hence the stochastic process and the profile of the spectral dimension) uniquely determined? If not, is the physics robust?

Here we give a qualitative anticipation of what will emerge during our analysis. This will help the reader in framing the salient characteristics of multiscale and multifractal geometries within the mathematical presentation, placing proper emphasis on those details that play a major role in the physics of quantum gravity.

Part of the introductory motivation was based on the observation that a spacetime dimension equal to 2 is universally associated with the UV finiteness of quantum gravity; dimensional flow from two limiting integer values (in particular, $\ds\sim 2$ to $\ds\sim 4$) is an almost universal feature of such theories. While the value 2 is itself universal, the IR limit $\sim 4$ is an empirical datum forced into the construction of the models, with barely one exception \cite{fra4}.

The number $N$ of asymptotic regimes in dimensional flow is not universal because there exist theories, or variations within the same theory, predicting more than two plateaux in the profile of $\ds(\ell)$, where $\ell$ is the probed length scale. Asymptotic safety (depending on the operators assumed in the action) and multifractional models (depending on their construction) are two instances. CDT show only two regimes but the small-scale one is probably an intermediate, not a deep UV feature of the model \cite{RSnax}.

Once $N$ is fixed, how the flow is realized is not universal either, inasmuch as different models with the same number of characteristic scales generate functionally different profiles $\ds(\ell)$. Thus, while the value of each plateau may be the same across all possible profiles and the interpolating regimes between them qualitatively similar, the actual function realizing the interpolation can be different. Nevertheless, we point out that a certain function $\ds(\ell)$ recurs often in the literature and can be regarded as general, albeit nonuniversal, for systems with just one characteristic scale $\ell_*$. In four topological dimensions, this function is
\be\label{gepr}
\ds(\ell)=4\frac{\a_*+(\ell/\ell_*)^2}{1+(\ell/\ell_*)^2}\,,
\ee
where typically $\a_*=1/2$ and, as we shall comment below, the power of the length ratio is of little importance. The instances where this profile is realized are at least the following.

\emph{(a) Multifractional spacetimes}. Because this framework aims to capture, by construction, universal or general multiscale geometries appearing in quantum gravity, the derivation of Eq.\ \Eq{prof1} (for isotropic measures) automatically contains an explanation of the generality of the profile \Eq{gepr} in single-scale scenarios. It simply stems from the most natural adaptation of rods in multiscale measurements.

\emph{(b) Dilaton black holes}, where Eq.\ \Eq{gepr} is an accurate estimate of an exact but more complicated profile $\ds(\ell)$ \cite{CaG}. Since the dilaton plays the role of a power-law measure weight in the action, technically this case is nothing but a radial version of a multifractional spacetime.

\emph{(c) Causal dynamical triangulations}. A numerical best fit of dimensional flow yields $\a_*\approx 1/2$ and $(\ell/\ell_*)^2\approx \s/50$, where $\s$ is the dimensionless diffusion time in the CDT transport equation \cite{AJL4}.

\emph{(d) Random multigraphs}, whose geometry was shown to be closely akin to (certain approximations of) causal dynamical triangulations \cite{GWZ1,GWZ2}. The analytic form of the return probability is, in fact, $\cP(\s)= 2G^2/(2G\s+\s^2)$, leading to
$\ds = 4[1/2+\s/(2G)]/[1+\s/(2G)]$, where $G$ is Newton's constant.

\emph{(e) Ho\v{r}ava--Lifshitz gravity}. There, the scaling of time and spatial coordinates is mutually anomalous and governed by a parameter $z$ which is fixed in the UV and in the IR (to the values 3 and 1, respectively). In general, $\ds=1+3/z$ \cite{Hor3}; this expression captures the asymptotic behavior of the spectral dimension in the two (UV and IR) plateaux corresponding to $z=3$ and $z=1$. The actual profile $\ds(\ell)$ would stem from the exact diffusion equation featuring (at least two) Laplacians of different order. The spatial generator of the diffusion equation can thus be written, effectively, as a sum over $z$ of Laplacians with $z$-dependent order \cite{SVW2}. In turn, $z$ can be regarded phenomenologically as a scale-dependent parameter, which reproduces Eq.\ \Eq{gepr} with $\a_*=1/2$ if $z(\ell)=[1+(\ell/\ell_*)^2]/[(1/3)+(\ell/\ell_*)^2]$.

The latter matching is justified by the evidence we shall gather that multiscale (in particular, multi-Laplacian) and multifractional diffusion equations yield the same spectral dimension profile when the scale hierarchy is mutually tuned.

\begin{figure}
\centering
\includegraphics[width=8.6cm]{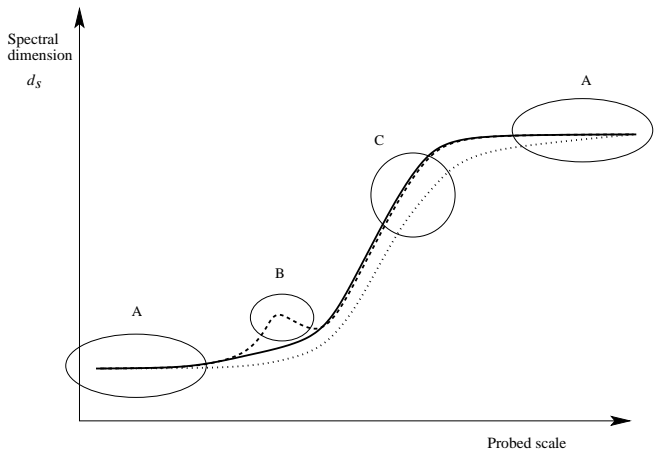}
\caption{\label{fig1} Typical single-scale profile of the spectral dimension as a function of the probed length scale. (\textit{A}) Asymptotic regimes where $\ds\sim {\rm const}$ (plateaux) and the values of $\ds$ therein are universal or almost universal. (\textit{B}) Also intermediate plateaux, possibly reduced to local extrema, are robust within a given physical system, but different mathematical realizations of the same system cannot produce extra plateaux or transient features such as bumps, glitches, and so on. (\textit{C}) Details of dimensional flow such as the monotonic slope of the profile between different regimes can change with the mathematical realization but they are physically unimportant.}
\end{figure}

If the number $N-1$ of characteristic or fundamental scales of the system is specified, so is the number $N\geq 2$ of asymptotic regimes in the spectral dimension. Profiles with different $N$ belong to physically different systems. Figure \ref{fig1} shows the typical two-regime (one characteristic scale, $N=2$) profile.

The question about uniqueness is more delicate and is summarized in the following double claim:
\begin{itemize}
\item Given a physical system, we can associate \emph{different} diffusion equations realizing very similar dimensional flows with the \emph{same} physical features, i.e., (i) the number $N-1$ of plateaux, (ii) their mutual position, (iii) their values, and (iv) the monotonicity of the flow between any two plateaux. 
\item Other features such as (v) the length of the plateaux and (vi) the way they are connected are \emph{not} unique, can vary depending on the mathematical implementation of the model, and are not expected to play any role as far as physical falsifiability is concerned.
\end{itemize}

The number $N$, the values of $\ds$ at the plateaux, and their mutual position are almost always default ingredients, specified by construction either explicitly (e.g., requiring that $\ds\sim 4$ in the IR, or introducing only $N-1=1$ fundamental scale such as the Planck scale in noncommutative geometries or the simplest multifractional setting) or implicitly (e.g., choosing the operators in the action in QEG determines $N$ and, in a rather mysterious way, the value and position of the intermediate plateau). This should highlight the intrinsically empirical nature of the concept of universality.

Throughout the paper a plethora of examples will show that, once $N$ and the values of the plateaux and their mutual positions are fixed, different diffusion equations will be able to generate an analytic dimensional flow connecting them in about the same way. Obviously, the profiles for $\ds$ will be different functions of the probed scale, but since $N$ is fixed there will be no intermediate local extrema (maxima or minima, ``bumps,'' ``glitches,'' or other transient features like the bump \textit{B} in Fig.\ \ref{fig1}) between two plateaux. Thus, universal or almost universal features are unique and robust, while the interconnecting monotonic slopes in the flow (feature \textit{C} in Fig.\ \ref{fig1}) are nonunique but robust.

In transport theory, it is not altogether unusual to associate different diffusion equations (i.e., different transport models) with the same physical system and, at the same time, being unable to place observational constraints on such models. The reason is that, in practice, only anomalous exponents can be determined by experiments, while details of the diffusion process are more elusive. The theory itself, however, can discriminate among the models, because diffusion equations with no probabilistic interpretation are much less appealing than those with a solution with stochastic meaning, even if they all realize the same anomalous behavior asymptotically. Below we will stress and illustrate this point on several occasions. 

In quantum gravity, where we have virtually no acknowledged observational signature available (with the possible exception of the cosmological constant), this limitation of ``uniqueness'' is all the more cogent. On top of that, there are explanations of why transient slopes in the dimensional flow are associated with nonunique mathematical features of the models: we refer, in particular, to regularization schemes in triangulation settings (as in CDT) and in renormalization group flows (as in asymptotic safety and Ho\v{r}ava--Lifshitz gravity). Therefore, even from the point of view of quantum field theory and simplicial geometry it is not always physically meaningful to expect transient regimes to be unique. Finally, we wish to advance the somewhat neglected idea that diffusion must have a well-defined stochastic interpretation in quantum gravity also. Diffusion, in fact, probes geometry itself via a test particle and one should be in a position to meaningfully ask what the probability is of finding the probe at a certain point.


\section{Geometry of classical fractional spacetimes}\label{cfsf}


\subsection{Configuration space and Hausdorff dimension}\label{f1}

Continuum models of fractal spacetimes are defined via an embedding, a measure $\vr$, and a Laplace--Beltrami operator $\cK$. Let $M^D$ be Minkowski spacetime in $D$ dimensions, labeled by Greek indices $\mu,\nu,\dots=0,1,\dots,D-1$. The measure is of a Lebesgue--Stieltjes type \cite{fra1,fra2} such that it can be recast as a Lebesgue measure with a nontrivial weight:
\be
\rmd\vr(x)=\rmd^Dx\,v(x)\,.
\ee
We assume that, in momentum units, the scaling dimension of the weight $v(x)$ is
\be
\begin{matrix}
& [v(x)]=-D(\a-1)\,,\qquad \a=\frac{1}{D}\sum_\mu \a_\mu\,,\\ & \\
& \qquad \frac12\leq\a_\mu\leq 1\,,\qquad \forall~\mu\,.\end{matrix}
\ee
Here the ``fractional charges'' $\a_\mu$ are parameters associated with each direction. An explicit realization of this weight is given by real-order fractional measures $\rmd\vr_\a(x)=\rmd^Dx\, v_\a(x)$, where \cite{frc1,frc3}
\be\label{frame}
v_\a(x)=\prod_\mu v_\a(x^\mu):=\prod_\mu \frac{|x^\mu|^{\a_\mu-1}}{\Gamma(\a_\mu)}\,,
\ee
where the absolute value is taken so that the measure is real-valued and positive. This has consequences for quantization \cite{frc4}, because unilateral measures with support only in the first orthant ($x^\mu\geq 0$) are eventually conjugate to problematic non-negative momenta. Hence, we take the integration range to be the whole real line for each coordinate. The ``isotropic case'' $\a_\mu=\a$ can be assumed, if desired, to illustrate the properties of these spaces. Real-order fractional measures make anomalous scaling apparent,
\be
\vr_\a(\la x)=\la^{\dh}\vr_\a(x)\,,\qquad \dh=D\a\,,
\ee
where $\dh$ is the Hausdorff dimension of spacetime. $\dh$ can be computed from a similarity theorem or from its operational definition, as the exponent in the power-law scaling of the volume of a $D$-ball with respect to the radius, $\cV^{(D)}(R)\sim R^{\dh}$ \cite{frc1}.


\subsection{Laplacians}\label{f2}

The so-called harmonic structure of the fractal Minkwoski spacetime $\cM_v^D$ is specified by the choice of Laplace--Beltrami operator. Consider, for instance, the action of a free massless scalar field:
\be\label{S}
S=\frac12\int_{-\infty}^{+\infty}\rmd\vr(x)\,\phi\cK\phi\,.
\ee
One can show that a natural Laplace--Beltrami operator on these spaces is given by the self-adjoint operator $\cD$:
\be
\cK=\eta^{\mu\nu}\cD_\mu\cD_\nu\,,\qquad \cD_\mu:=\frac{1}{\sqrt{v(x)}}\,\p_\mu\left[\sqrt{v(x)}\,\,\cdot\,\right]\,,
\ee
where $\eta_{\mu\nu}$ is the Minkowski metric with signature $(-,+,\cdots,+)$, Einstein summation convention is used, and $\p_\mu=\p/\p x^\mu$ is the ordinary partial derivative. $\cK$ is a second-order differential operator guaranteeing that, upon integrating by parts, the kinetic term in \Eq{S} can also be written in the symmetric form $-\cD_\mu\phi\cD^\mu\phi$ (boundary terms vanish on a suitable space of functionals \cite{frc3}).

These properties are highly desirable, or even necessary, when considering fractal field theory as fundamental. As an effective model, however, $\cM_v^D$ can be equipped with more general non-Hermitian Laplacians. Consider, in one dimension, the left and right Caputo derivatives
\ba
({}_a\p^\g f)(x) &:=&  \frac{1}{\Gamma(n-\g)}\int_{a}^{x} \frac{\rmd x'}{(x-x')^{\g+1-n}}\p^n_{x'}f(x')\,,\nonumber\\
&& n-1\leq \g<n\,,\qquad x>a\,,\label{pan}\\
({}_b\bp^\g f)(x) &:=& \frac{(-1)^n}{\Gamma(n-\g)}\int_{x}^{b} \frac{\rmd x'}{(x'-x)^{\g+1-n}}\p^n_{x'}f(x')\,,\nonumber\\ 
&& n-1\leq \g<n\,,\qquad x<b\,,\label{bpan}
\ea
where $\p$ is the ordinary first-order partial derivative and $n\geq 1$ is a natural number. A review of the properties of Caputo integro-differential operators can be found in Ref.\ \cite{frc1} and references therein. Here we only recall some features. When $\g\to n$, ${}_a\p^{n}=\p^{n}=(-1)^{n}{}_b\bp^{n}$. 
Left and right derivatives are mutually related by a reflection $x\to a+b-x$. In our context, the fractional measure splits the real line in two at $x=0$, so that the natural fractional derivatives are $\{{}_{-\infty}\p^\g, {}_{+\infty}\bp^\g\}$ (also called the Liouville and Weyl derivative, respectively) and $\{{}_{0}\p^\g, {}_{0}\bp^\g\}$, which we shall denote, respectively, as
\be
\{{}_{\infty}\p^\g, {}_{\infty}\bp^\g\}\,,\qquad \{\p^\g, \bp^\g\}\,.
\ee
We paired derivatives which are conjugate under reflection, in both cases at 0, $x\to -x$. By conjugate, we mean
\bs\ba
&&(\p^\g f)(x) = (\bp^\g f_-)(-x)\,,\nonumber\\
&&\qquad (\bp^\g f)(x)=(\p^\g f_-)(-x)\,,\label{pari3a}\\
&& ({}_{\infty}\p^\g f)(x) = ({}_{\infty}\bp^\g f_-)(-x)\,,\nonumber\\
&&\qquad ({}_{\infty}\bp^\g f)(x)=({}_{\infty}\p^\g f_-)(-x)\,,\label{pari3b}
\ea\es
where $f_-(x):=f(-x)$. While for ordinary derivatives this parity transformation leads to the same operator up to a sign, $\p \to -\p$, in fractional calculus it transforms left into right derivatives, and vice versa. For instance \cite[Eq.~(2.3.23)]{KST},
\be\label{ibp}
\int_{-\infty}^{+\infty}\rmd x\, f\,{}_{\infty}\p^\g g = \int_{-\infty}^{+\infty}\rmd x\, g\,{}_{\infty}\bp^\g f\,.
\ee
This is the reason why one cannot construct naive self-adjoint fractional Laplacians on fractional spaces. In Ref.\ \cite{frc1}, the fractional operators
\be\nonumber
\cK_\g  = \eta^{\mu\nu}\p_\mu^\g\p^\g_\nu\,,\qquad \bar\cK_\g = \eta^{\mu\nu}{}_\infty\bp_\mu^\g\,{}_\infty\bp^\g_\nu
\ee
were defined for a unilateral world where $x^\mu\geq 0$ (a subscript $\mu$ denotes the left or right Caputo derivative with respect to $x^\mu$). However, in a bilateral world the coordinate upon which one differentiates must lie within the integration range of the given derivative, so one can use both ${}_{\infty}\p^\g$ and ${}_{\infty}\bp^\g$ at the same time, or else the pair $\bp^\g$ and $\p^\g$. Therefore, adding some weight factors for later convenience, here we shall define the fractional Laplace--Beltrami operators\footnote{We call them ``fractional'' even if in many formul\ae\ the measure $v(x)$ may be more general than a fractional measure $v_\a(x)$.}
\ba
\bar\cK_{\g,\a}     &:=& \frac{1}{\sqrt{v(x)}}\,\eta^{\mu\nu}\left(c_\g\,\p_\mu^\g\p^\g_\nu+\bar c_\g\,\bp_\mu^\g\bp^\g_\nu\right)\left[\sqrt{v(x)}\,\,\cdot\,\right]\,,\nonumber\\\label{cka1}\\
\cK_{\g,\a} &:=& \frac{1}{\sqrt{v(x)}}\,\eta^{\mu\nu}\nonumber\\
&&\times\left(c_\g\,{}_\infty\p_\mu^\g{}_\infty\p^\g_\nu+\bar c_\g\,{}_\infty\bp_\mu^\g{}_\infty\bp^\g_\nu\right)\left[\sqrt{v(x)}\,\,\cdot\,\right],\label{cka2}
\ea
where $c_\g$ and $\bar c_\g$ are complex coefficients dependent on $\g$, which we choose so that $c_n+\bar c_n=1$ for $n$ natural. Suitable boundary conditions at $x=0$ on the space of functionals yield the properties $\p^\g\p^\g=\p^{2\g}$ and $\bp^\g\bp^\g=\bp^{2\g}$, while for Liouville derivatives it is always true that ${}_{\infty}\p^\g{}_{\infty}\p^\g={}_{\infty}\p^{2\g}$, ${}_{\infty}\bp^\g{}_{\infty}\bp^\g={}_{\infty}\bp^{2\g}$ \cite[Sec.\ 2.3.2]{frc1}. Thus, the order of the Laplace--Beltrami operators \Eq{cka1} and \Eq{cka2} is $2\g$, where $\g=\sum_\mu\g_\mu/D$. In the isotropic case, $\g_\mu=\g$ for all $\mu$. In anisotropic formulations, summation with the Minkowski metric must be accompanied by suitable dimensionful coefficients, which we omitted in Eqs.~\Eq{cka1} and \Eq{cka2}. When $\g=1$, $\bar\cK_{1,\a}=\cK=\cK_{1,\a}$; when $\g=n$ and $\a=1$, $\bar\cK_{n,1}=\cK_{n,1}=\eta^{\mu\nu}\p_\mu^n\p_\nu^n$. In Euclidean signature, we denote the ordinary Laplacian as
\be\label{lapla}
\N^2 = \sum_{\mu=1}^D \frac{\p^2}{\p x_\mu^2}\,,
\ee
and, in general, higher-order differential operators as
\be\label{laplan}
\N^n = \sum_{\mu=1}^D \frac{\p^n}{\p x_\mu^n}\,,\qquad n\in\mathbb{N}\,.
\ee
With the measure \Eq{frame} and bilateral integration, from Eq.~\Eq{ibp} one sees that the operator \Eq{cka2} (with coefficients chosen as below) is self-adjoint with respect to the scalar product of the Hilbert space of suitably ``good'' functions given by the intersection of the domain of $\cK_{\g,\a}$ and the space of functions \cite{frc3} which are $L^2$ with respect to the measure $\vr$. Given two such nontrivial functions $f$ and $g$, the scalar product is defined as
\be
(f\,,\,g) := \int_{-\infty}^{+\infty}\rmd\vr(x)\,f^*(x)\,g(x)\,,
\ee
where the asterisk indicates complex conjugation, so that
\be\label{scala}
(f\,,\,\cK_{\g,\a} g) = (\cK_{\g,\a}f\,,\, g)\qquad\Leftrightarrow\qquad \bar c_\g=c_\g^*\,.
\ee
From now on, we set $\bar c_\g=c_\g^*$ in the definition \Eq{cka2}. This argument is sufficient to exclude $\bar\cK_{\g,\a}$ (which cannot be Hermitian on this space) but we shall give another motivation soon.

In a field-theory setting, combinations of Laplacians with integer $\g\geq 1$ will, in general, introduce ghost modes (e.g., \cite{Smi04,BMS,BaK1}), while fractional Laplacians ($0<\g\notin\mathbb{N}$) are typically associated with continuum quasiparticle spectra. The first fact already foretells an issue with values $\g\geq 1$ which will be further strengthened when looking at probabilities in diffusion processes.

For later use, we recall that (e.g., \cite{frc1})
\ba
\p^\g E_\g(\la x^\g)           &=& \la E_\g(\la x^\g)\,,\label{mile}\\
\p^\g             \rme^{\la x} &=& \la x^{1-\g}E_{1,2-\g}(\la x)\,,\label{espo1}
\ea
where $E$ is the Mittag-Leffler function
\bs\label{mil}\ba
E_a(z) &:=& E_{a,1}(z)\,,\\
E_{a,b}(z)&:=&\sum_{n=0}^{+\infty}\frac{z^n}{\Gamma(an+b)}\,,\qquad a>0\,.
\ea\es
Analogous expressions hold for the right derivative with $x\to -x$. Also,
\be
{}_{\infty}\p^\g \rme^{\la x} = \la^\g\rme^{\la x}\,,\qquad {}_{\infty}\bp^\g \rme^{\la x} = (-\la)^\g\rme^{\la x}\,.\label{espo2}
\ee


\subsection{Momentum transform}\label{f3}

We restrict now the measure to be non-negative definite and factorizable with respect to its coordinate dependence:
\be
v(x)=\prod_\mu v_\mu(x^\mu)\,,\qquad v_\mu(x^\mu)\geq 0\,,\qquad \forall~ x^\mu\,,
\ee
where the index $\mu$ in $v_\mu$ is not vectorial. The fractional measure \Eq{frame} satisfies these conditions. Let $\cM_w^D$ be another space in a $D$-dimensional embedding, spanned by ``momenta'' $k$ and with measure
\be
\rmd\tau(k)=\rmd^Dk\,w(k)\,,\qquad \dh=D\a'\,,
\ee
where $\a'$ is not necessarily equal to $\a$. If $w(k)$ obeys the same properties as $v$ [as in the case of two fractional measures $v=v_\a(x)$ and $w=v_{\a'}(k)$], then there exists a family of unitary invertible mappings $F:\cM_v^D\to \cM_w^D$ identifying $\cM_w^D$ as momentum space \cite{frc3}. This is the generalization of the Fourier transform on Lebesgue--Stieltjes spacetimes with factorizable measures:
\bs\label{fbit12}\ba
\tilde f(k) &:=& \int_{-\infty}^{+\infty}\rmd\vr(x)\,f(x)\,\bE^*(k,x)=:F_{\vr,\tau}[f(x)],\label{fo1}\\
f(x) &=& \int_{-\infty}^{+\infty}\rmd\tau(k)\,\tilde f(k)\,\bE(k,x)\,,\label{fo2}
\ea\es
where $\bE(k,x)$ are the kernel functions on which the transform is expanded. By construction, they give an integral representation of the fractal Dirac distributions in position and momentum space:
\ba
\int_{-\infty}^{+\infty}\rmd\tau(k)\,\bE^*(k,x)\bE(k,x') &=& \de_v(x,x')\nonumber\\
&:=&\frac{\de(x-x')}{\sqrt{v(x)v(x')}}\,,\label{KK}\\
\int_{-\infty}^{+\infty}\rmd\vr(x)\,\bE^*(k,x)\bE(k',x) &=& \de_w(k,k')\nonumber\\
&:=&\frac{\de(k-k')}{\sqrt{w(k)w(k')}}\,,
\ea
which are quite generally nontranslationally invariant. When $v=w$, $F_{\vr,\tau}$ is an automorphism, the $\delta$ distribution is the same in both spaces and the kernels $\bE$ are symmetric in $x$ and $k$. 


\subsection{Eigenvalue equations}\label{f2b}

In Ref.\ \cite{frc3}, it was shown that the $\bE$'s are either weighted Bessel functions of the first kind or weighted phases, where the weight is $[w(k)v(x)]^{-1/2}$. For calculational purposes, we demand that the kernel functions $\bE$ are eigenfunctions of the chosen Laplacian operators. However, from Eq.~\Eq{mile} one sees that none of the kernels can be eigenfunctions of the operator $\bar\cK_{\g,\a}$, so we cannot use the momentum transforms to diagonalize this Laplacian. From now on we drop $\bar\cK_{\g,\a}$ from the discussion. On the other hand, phases are eigenfunctions of the ordinary derivative, of the left derivative from $-\infty$ to $x$ and of the right derivative from $x$ to $+\infty$ [Eq.~\Eq{espo2}]. In particular, the phaselike kernel 
\be\label{kern}
\bE(k,x) = \frac{1}{\sqrt{w(k)v(x)}}\,\frac{\rme^{\rmi k\cdot x}}{(2\pi)^{\frac{D}{2}}}
\ee
is an eigenfunction of both $\cK$ and $\cK_{\g,\a}$:
\ba
\cK\,\bE(k,x)        &=& -k^2 \bE(k,x)\,,\\
k^2&:=& k_\mu k^\mu=-k_0^2+k_1^2+\dots+k_{D-1}^2\,,\nonumber\\
\cK_{\g,\a}\,\bE(k,x) &=& F^{2\g}(k)\, \bE(k,x)\,.\label{eiK}
\ea
We now determine the form of the eigenvalue $F^{2\g}(k)=F^{2\g}(k^0,k^1,\dots,k^{D-1})$. Take a function $f(x)=h(kx)$, where $k\neq 0$ is a constant (later one can analytically continue the final result to $k=0$). From Eq.~\Eq{bpan} one finds that
\ba
({}_\infty\p^{2\g}_x f)(x) &=& ({}_\infty\p^{2\g}_x h)(kx)\nonumber\\
&\stackrel{z=kx}{=}& \frac{1}{\Gamma(n-2\g)}\int_{-{\rm sgn}(k)\infty}^{z}\rmd z'\nonumber\\
&&\times\frac{k^{n-1}}{[k^{-1}(z-z')]^{2\g+1-n}}\,\p^n_{z'}h(z')\nonumber\\
&=& k^{2\g}\theta(k)\,({}_{\infty}\p^{2\g}_z h)(z)\nonumber\\
&&+(-k)^{2\g}\theta(-k)\,({}_{\infty}\bp^{2\g}_z h)(z)\nonumber\\
&=& |k|^{2\g}[\theta(k)\,({}_{\infty}\p^{2\g}_z h)(z)\nonumber\\
&&+\theta(-k)\,({}_{\infty}\bp^{2\g}_z h)(z)]\,,\label{use1}
\ea
where $\theta$ is the Heaviside distribution:
\be
\theta(k)=\left\{ \begin{matrix} 1\,,&\quad k> 0\,,\\
                                 0\,,&\quad k<0\,.\end{matrix}\right. 
\ee
In what follows, the value of $\theta$ at $k=0$ is irrelevant as $\g>0$. Similarly, for the right Liouville derivative
\ba
({}_\infty\bp^{2\g}_x f)(x) &=& |k|^{2\g}[\theta(k)\,({}_{\infty}\bp^{2\g}_z h)(z)\nonumber\\
&&+\theta(-k)\,({}_{\infty}\p^{2\g}_z h)(z)]\,,\label{use2}
\ea
so that we can combine Eqs.~\Eq{use1} and \Eq{use2} into the operator
\ba
(\Delta_\g f)(x)&:=& c_\g\,({}_\infty\p^{2\g}_x f)(x)+\bar c_\g\,({}_\infty\bp^{2\g}_x f)(x)\nonumber\\ &=&|k|^{2\g}[c_\g(k)\,({}_{\infty}\p^{2\g}_z h)(z)\nonumber\\
&&\quad+c_\g(-k)\,({}_{\infty}\bp^{2\g}_z h)(z)]\,,
\ea
where $c_\g(k):=\theta(k)\,c_\g+\theta(-k)\,\bar c_\g$. 

For the function $f(x)=h(kx)=\rme^{\la k x}$, we get
\be
\Delta_\g \rme^{\la k x}=[c_\g(k)\,\la^{2\g}+c_\g(-k)\,(-\la)^{2\g}]|k|^{2\g}\rme^{\la k x}\,,
\ee
so that, for $\la=\pm\rmi=\rme^{\pm\rmi\pi/2}$, we have
\ba
\Delta_\g \rme^{\pm\rmi kx}&=&[c_\g(k)\,\rme^{\pm\rmi\pi\g}+c_\g(-k)\,\rme^{\mp\rmi\pi\g}]|k|^{2\g}\rme^{\pm\rmi kx}\nonumber\\
&=:&c_{{\rm sgn}(k)\g}^\pm\,|k|^{2\g}\rme^{\pm\rmi kx}\,.\label{phases}
\ea
Some remarks about the eigenvalues of $\Delta_\g$ follow:
\begin{itemize}
\item[(a)] Choosing $\bar c_\g\neq c_\g^*$ can lead to complex-valued spectra.
\item[(b)] Otherwise, for $\bar c_\g= c_\g^*$ the spectrum of $\Delta_\g$ has a discontinuity at $k=0$ and, thanks to the self-adjointness of the Laplacian, it is real. Writing $c_\g= |c_\g|\,\rme^{\rmi\pi\vp}$, one has
\be
c_{{\rm sgn}(k)\g}^\pm=c_\g^\pm := 2|c_\g|\cos[\pi(\vp\pm\g)]\,.
\ee
\item[(c)] If $c_\g=1/(2\rmi)$ ($\vp=-1/2$), then $c^\pm_\g=\pm\,{\rm sgn}(k)\,\sin(\pi\g)$ and the signature of the spectrum depends on the sign of both $k$ and the phase.
\item[(d)] If $c_\g=1/2$ ($\vp=0$), then $c^\pm_\g=\cos(\pi\g)$ depends neither on the sign of $k$ nor on that of the phase, and the spectrum is semibounded. When $\g=n+1/2$ is a half integer, $n\in\mathbb{Z}$, plane waves are in the kernel of $\Delta_\g$. When $\g=n$, $c^\pm_\g=(-1)^{n}$ and the spectrum is positive (negative) definite for $n$ even (odd), with zero eigenvalue at $k=0$.
\end{itemize}
Moving to many dimensions and the operator $\cK_{\g,\a}$, we conclude that the eigenvalue in Eq.~\Eq{eiK} is
\be
F^{2\g}(k) := -c_{{\rm sgn}(k^0)\g}^+\,|k^0|^{2\g}+\sum_{i=1}^{D-1} c_{{\rm sgn}(k^i)\g}^+\,|k^i|^{2\g}\,.
\ee
Since the sign of the phase should not play any role, we will often choose the coefficients
\be
c_\g=\frac{1}{2}\,\rme^{\rmi\pi(1-\g)}\,,\qquad F^{2\g}(k) := |k^0|^{2\g}-\sum_{i=1}^{D-1} |k^i|^{2\g}\,.
\ee

Other operators are possible whose eigenvalues have fractional momentum dimension. Rotation-invariant examples are Riesz Laplacian \cite[Secs.\ 2.10 and 5.5]{KST} and the Riesz--Bessel operator \cite{ARAG,AAMR}, and, in Lorentzian signature, the Lorentz-covariant fractional powers of the d'Alembertian, $(\B-m^2)^\g$, corresponding to a spectrum $F^{2\g}(k)=-(k^2+m^2)^\g$ \cite{BGG,Gia91,BGO,doA92,BG,BOR,BBOR}.


\section{Diffusion equations and stochastic processes}\label{f4}

The spectral dimension of $\cM_v^D$ is defined via a diffusion process \cite{frc1}. Here we extend that calculation to the case of bilateral measures and fractional Laplacians, as well as to various classes of stochastic processes associated with different ranges in the diffusion exponents $\b$ and $\g$. We review these processes from the literature of stochastic theory and comment on the physical meaning of the heat kernel as diffusion probability. 

To simplify the logic presentation, it is convenient to specialize first to the case of fixed dimensionality (no scale hierarchy). We concentrate on the continuum formulation of fractional calculus, which guarantees anomalous (in particular, fractal) geometric properties of spacetimes \cite{frc1,frc2,frc3} and anomalous correlations in diffusion problems (e.g., \cite{MeK,Zas3,MeK2,MCK,Sok12}). Another assumption we make here is to ignore curvature. The latter modifies the spectral properties of spacetime even in a classical setting, except in the UV limit $\s\to 0$. In the absence of dimensional flow, one can include this limit in the definition of the spectral dimension, but in general this would prevent the probing of large scales. Effective quantum spacetimes, anyway, are modified by quantum geometric effects even in the absence of curvature, both globally and locally in inertial frames, which motivates the assumption.\footnote{Another point of view is that a multiscale geometry is very often probed in ``snapshots'' taken at different given scales; these snapshots are then collected to give a fragmentary picture of dimensional flow, typically in the UV and IR limits and in transient regimes. Each snapshot is related to a fixed spectral dimension and the limit $\s\to 0$ is legitimate. In our approach we see global modifications of spacetime \emph{and} we have full control of dimensional flow, so we do not adopt this perspective.}

As in the ordinary case \Eq{dife}, the geometry is probed by a process governed by a hyperbolic diffusion equation (or fractional wave equation, for $1<\b\leq 2$), but of the form
\bs\label{difef}\ba
&& (\p_\s^\b-\cK_{\g,\a}^{\rm E})P(x,x',\s)=\cS(x,x',\s)\,,\\
&&  P(x,x',0)=\delta_v(x,x')\,,
\ea\es
where $\p^\b_\s$ is the Caputo diffusion operator (we do not consider ${}_\infty\bp^\b_\s$ since $\p_\s^\b$ is the natural derivative for a process with memory loss starting at $\s=0$), $\cK_{\g,\a}^{\rm E}$ is the Euclideanized version of Eq.~\Eq{cka2}, $P(x,x',\s)$ is the heat kernel and $\cS$ is a source term. The external ``time'' or ``scale'' $\s$ has actually dimension $[\s]=-2\g/\b$. As far as the author knows, the source $\cS$ has been set to zero in all diffusion processes in quantum gravity considered in the literature.

Different ranges in the parameters $\b$ and $\g$ can correspond to physically inequivalent diffusion processes, for which there exists extensive literature (for $\a=1$ and \emph{other} types of fractional Laplacians) \cite{MeK,Zas3,MeK2,MCK,MLP}. Set, first, $\a=1$ and $\cS=0$.
\begin{itemize}
\item[(a)] When $\b=\g=1$ (Eq.\ \Eq{dife}), one has ordinary diffusion with the solution
\be\label{gau}
P(x,x',\s)=u_1(r,\s):=\frac{\rme^{-\frac{r^2}{4\s}}}{(4\pi\s)^{\frac{D}{2}}}\,,
\ee
where $r^2:=\sum_\mu|x_\mu-x_\mu'|^2$ is the Euclidean distance between the two points. The diffusion equation is associated with a \emph{Wiener process} $B(\s)$, also known as \emph{standard Brownian motion}. A Wiener process is such that (i) $B$ is continuous in $\s$ almost surely (i.e., with probability 1), (ii) $B(0)=x'$, and (iii) the increments of $B$ are independent and governed by the Gaussian distribution $u_1$, so that $B(\s)-B(\s')\sim u_1(0,\s-\s')$ for $\s'<\s$. Here $B$ is the random variable denoting the position $x$ of the particle at time $\s$ (in $D$ dimensions, it is a vector).
\item[(b)] When $0<\b<1$ and $\g=1$, the process possesses a heavy tail in waiting times, leading to a delay of particle diffusion and, hence, to \emph{subdiffusion}.\footnote{The fractional-time diffusion equation is not associated with fractional Brownian motion \cite{Sok12,Kol40,MaV}, whose probability density function is different from the solutions in Sec.\ \ref{ftde}.}
\item[(c)] When $\b=1$ and $0<\g<1$, the process is a \emph{L\'evy process}. A L\'evy process $L$ has the following properties: (i) $L(\s)$ is right continuous with left limits almost surely, (ii) $L(0)=x'$ almost surely, (iii) for any sequence $\s_{n-1}<\s_n$, the increments $L(\s_n)-L(\s_{n-1})$ are independent, and (iv) their distribution is equal to the distribution of $L(\s_n-\s_{n-1})$. It is a self-similar process like fractional Brownian motion, but with an island-type structure, characterized by a heavy-tailed distribution and ``long jumps'' connecting clusters of shorter steps. This heavy tail in space is associated with \emph{superdiffusion}.
\item[(d)] Systems outside the range
\be\label{0bg1}
0<\b,\g\leq 1
\ee
correspond to transport pseudoprocesses for which Eq.~\Eq{difef} may not provide a well-defined diffusion model in a probabilistic sense (i.e., the signature of $P$ may be indefinite, depending on the case and on the presence of a source $\cS\neq 0$). Probabilistic processes do exist for $1<\b\leq 2$ when $\g=1$ or $\g\geq\b/2$ in one dimension \cite{MLP}. Models of other systems are still unknown, but there exists a wealth of studies on pseudoprocesses.
\end{itemize}
For $\a\neq 1$, one introduces extra friction and potential terms. When $\b=\g=1$, diffusion is said to be \emph{normal} or Gaussian (the mean squared displacement of the test particle grows linearly with diffusion time), otherwise it is \emph{anomalous}.

In general, processes described by a fractional or higher-order diffusion equation are not Markovian, meaning that, in contrast to Markov processes, future states depend on the present state but also on past ones, and the Chapman--Kolmogorov equation does not hold. This feature is expected in the case of fractional diffusion equations, where memory drag is typical of pseudodifferential operators. However, it holds also for integer higher-order equations where the initial condition appears explicitly as a source term.


\subsection{Solution scheme and probabilistic interpretation for $\cS=0$}\label{s0}

To solve the diffusion equation, assume $v(x)$ is factorizable and write $P$ and $\cS$ in terms of the kernel functions,
\ba
P(x,x',\s) &=& \int_{-\infty}^{+\infty}\rmd\tau(k)\,f_k(\s)\,\bE(k,x)\bE^*(k,x')\,,\label{diffkernel}\\
\cS(x,x',\s) &=& \int_{-\infty}^{+\infty}\rmd\tau(k)\,s_k(\s)\,\bE(k,x)\bE^*(k,x')\,,
\ea
where the function $f_k$ must solve the fractional differential equation
\be\label{gendiff}
[\p_\s^\b-F^{2\g}(k)]f_k(\s)= s_k(\s)\,,\qquad f_k(0)=1\,.
\ee
Notice that the initial condition in Eq.\ \Eq{difef} is respected, by virtue of Eq.~\Eq{KK}. 

Setting $s_k=0$ for simplicity, the solution of Eq.\ \Eq{gendiff} is
\be
f_k(\s)=\left\{\begin{matrix} \rme^{\s F^{2\g}(k)}\,,           &\qquad&  \b=1\,,\\
                              E_\b[F^{2\g}(k)\s^\b]\,,          &\qquad&  \text{$\b$ general}\,,\end{matrix}\right.\label{kkk}
\ee
where 
\be\label{eila}
F^{2\g}(k)=-\sum_\mu |k^\mu|^{2\g}
\ee
is the Euclidean eigenvalue of the Laplacian with the ``Riesz choice.'' Notice that, for $\g=\a$, $-F(k)=|k|$ is the $2\a$-norm of $k^\mu$, naturally equipping fractional spaces \cite{frc1}. Explicity, Eq.~\Eq{diffkernel} reads
\ba
P(x,x',\s) &=& \frac{1}{\sqrt{v(x)v(x')}}u_\b(x-x',\s)\,,\label{diffkernel2}\\
u_\b(x-x',\s) &:=& \int_{-\infty}^{+\infty}\frac{\rmd^Dk}{(2\pi)^D}\,f_k(\s)\,\rme^{\rmi k\cdot (x'-x)}\,.
\ea
For later convenience, we make the $\b$ dependence explicit via a subscript in $u$. The measure prefactor shows that $P$ is not translationally invariant. 

The full calculation of $P$ is not actually needed to get the spectral dimension, but it is extremely useful for clarifying the physical meaning of the diffusion equation. When the Riesz Laplacian is chosen, the integral can be done exactly, due to the fact that $f_k$ is rotationally invariant. However, for the fractional Laplacian \Eq{cka2} with eigenvalues \Eq{eila} the solution of the multidimensional fractional diffusion equation is considerably more demanding than in the one-dimensional case.\footnote{Its form is akin, but not equal, to the heat kernel in $D$ dimensions, for which several formulations and techniques (exact, asymptotic, and numerical) have been developed. For simplicity we consider the case where the fractional order $\g$ and the fractional charge $\a$ are the same in all directions, although anisotropic configurations $\g_\mu\neq\g_\nu$ can be of much interest \cite{MBB1,MBB2,WeZ1,SBMB,ZBMLS,QiH}. Anisotropic transport is, in fact, typical of Hamiltonian systems with nonergodic dynamics and a hierarchical set of islands, where particle jump distributions are asymmetric. Examples in nature are porous media in geologic aquifers \cite{RaG,ZhW}.} Here, we give an incomplete discussion of the properties of $P$. First, we argue that, in $D$ dimensions, $P$ is not positive definite for $\g>1$. Then we find the exact solution and its asymptotic behavior in $D=1$, and then in $D$ generic for $\b=1$ or $\g=1$. Many of the steps are adaptations of their analogs in ordinary spacetimes to the case of a fractional texture.\footnote{A possible source of confusion might be that the fractional diffusion equation with $\a=1$, i.e., the one considered in the literature of chaos theory and percolation systems, already describes a diffusive process on a fractal structure. Analogously, in quantum gravity one could interpret the fractional wave equation with $\a=1$ as a diffusion equation on a fractal spacetime; this is the case of QEG, for instance, where $\a=1$ and spacetime does have fractal properties. The introduction of a nontrivial measure weight $v(x)$ in position space further changes the background geometry and topology, as happened in some first attempts to generalize the diffusion equation to fractals \cite{HeP,OSP,MGN}.}

For general $\b$ and $\g$, $f_k$ is the Mittag-Leffler function of order $\b$. Equation \Eq{diffkernel2} implies the normalization
\bs\label{norp}\ba
&&\int_{-\infty}^{+\infty}\rmd^D x\,v(x)\,\sqrt{\frac{v(x')}{v(x)}} P(x,x',\s)\nonumber\\
&&\quad=\int_{-\infty}^{+\infty}\rmd^D x\,\sqrt{v(x)v(x')}\,P(x,x',\s)\\
&&=\int_{-\infty}^{+\infty}\rmd^D x\,u_\b(x-x',\s)=1\,,
\ea\es
where we chose the natural symmetrization in $x$ and $x'$. The extra factor in Eq.\ \Eq{norp} suggests to consider $\tilde P(x,x',\s)=\sqrt{v(x')/v(x)} P(x,x',\s)=u_\b(x-x',\s)/v(x)$ as the probability density defining the fractional system. The extra factor $\sqrt{v(x')/v(x)}$ would leave both the initial condition and the return probability unchanged (it is 1, respectively, on the support of the delta and when $x=x'$) but it would imply that the Laplacian in Eq.\ \Eq{difef} does not correspond to the self-adjoint operator appearing in the dynamics. In particular, it should be of the form $\check{\cK}=v^{-1}(x)\p_\mu\p^\mu [v(x)\,\cdot\,]$. This can be justified by unravelling the natural structure of the probability density as a fractional bilinear; this point will be further discussed elsewhere.

If we interpret the heat kernel as the probability to find the particle at $x$ at time $\s$ with initial position $x'$ at time $\s=0$, then we must require that $u_\b\geq 0$. Consequently, we can define the moments along the direction $\mu$ as
\ba
\langle |x-x'|^a\rangle_\mu &:=& \int_{-\infty}^{+\infty}\rmd^D x\,u_\b(x-x',\s)\,|x_\mu-x_\mu'|^a\,,\nonumber\\
&&\quad a>0\,,\label{mome}
\ea
where we expressed the integral already in terms of $u_\b$ since measure factors cancel one another. The definition of the moments allows us to show that $u_\b$ (hence $P$ and $\tilde P$) is not positive definite for $\g>1$. In general, not all moments will be finite but we assume that $\langle |x-x'|^2\rangle_\mu<\infty$. This is the case in $D=1$ and for all the $D$-dimensional special cases below; it is quite reasonable to expect finiteness of the second moment in general, lest the walk dimension be ill defined. Assuming also that $u_\b\geq 0$, we have $\langle |x-x'|^2\rangle_\mu >0$. Transforming $u_\b$ to momentum space,
\be
f_k(\s) = \int_{-\infty}^{+\infty}\rmd^D x \,u_\b(x-x',\s)\,\rme^{\rmi k\cdot (x-x')}\,,
\ee
one also notices that
\ba
\langle |x-x'|^2\rangle_\mu &=& -\frac{\p^2}{\p k_\mu^2} \left[f_k(\s)\big|_{k_\mu\neq k_\nu=0}\right]\Big|_{k_\mu=0}\nonumber\\
 &=& \sum_{n=1}^{+\infty} (-1)^{n+1} \frac{2\g n(2\g n-1)\s^{\b n}}{\Gamma(\b n+1)}\nonumber\\
 &&\times k_\mu^{2(\g n-1)}\Big|_{k_\mu=0}\,.\nonumber
\ea
This expression identically vanishes if $\g>1$, which contradicts the assumptions. Therefore, $P$ is not positive definite if $\g>1$. This conclusion holds only if $\cS=0$ and it cannot say anything about cases where the diffusion equation has a source.


\subsection{General solution in one dimension for $\cS=0$}\label{s1}

In $D=1$ dimension, $\a=1$ and in the absence of source, one can find the exact solution of
\be\label{1dde}
(\p_\s^\b-\p^{2\g}_{|x|})\,u_\b(x-x',\s)=0
\ee
and prove that $u_\b$ is non-negative definite for the range \Eq{0bg1}. The solution is the same as for the Riesz Laplacian (e.g., \cite{Zas3,MLP}), and in fact we denoted as $\p^{2\g}_{|x|}$ the Riesz-type spatial generator, i.e., the operator $\cK_{\g,1}$ in one dimension. The integral to solve is the Fourier transform of the Mittag-Leffler function. Rescaling the coordinates to the dimensionless variables
\be\label{reskx}
\tilde k =\s^{\frac{\b}{2\g}} k\,,\qquad \tilde x = \s^{-\frac{\b}{2\g}}x\,,
\ee
we get
\ba
&&\int_{-\infty}^{+\infty}\frac{\rmd k}{2\pi}\,E_\b(-|k|^{2\g}\s^\b)\,\rme^{\rmi k(x'-x)}\nonumber\\
&&\qquad\qquad \stackrel{\Eq{mil}}{=}\ \s^{-\frac{\b}{2\g}}\sum_{n=0}^{+\infty}\frac{(-1)^n}{\Gamma(\b n+1)}\nonumber\\
&&\times\int_{-\infty}^{+\infty}\frac{\rmd \tilde k}{2\pi} \,|\tilde k|^{2\g n}\rme^{\rmi \tilde k(\tilde x-\tilde x')}\nonumber\\
&&\qquad\qquad~ = \s^{-\frac{\b}{2\g}}\sum_{n=0}^{+\infty}\frac{(-1)^n}{\pi |\tilde x-\tilde x'|^{2\g n+1}}\frac{\Gamma(2\g n+1)}{\Gamma(\b n+1)}\nonumber\\
&&\qquad\qquad\qquad \times\cos\left[\frac{\pi}{2}(2\g n+1)\right]\,,\nonumber
\ea
where we used formula 17.21.25 of Ref.\ \cite{GR}, valid for noninteger $\g$. The series is absolutely convergent if $2\g<\b$ and can be resummed exactly for $\b=2\g$ and also for certain values outside this range \cite{SZ}. The (translation-invariant part of the) heat kernel is thus
\ba
u_\b(x-x',\s) &=&\s^{-\frac{\b}{2\g}}\sum_{n=0}^{+\infty}\left[(-1)^{n+1}\frac{\sin(\pi\g n)}{\pi}\frac{\Gamma(2\g n+1)}{\Gamma(\b n+1)}\right]\nonumber\\
&&\times \frac{1}{|\tilde x-\tilde x'|^{2\g n+1}}\,.\label{hek1}
\ea
Positiveness of $u_\b$ (and, hence, of $P$) is guaranteed if the fractional exponents are in the range \Eq{0bg1} (in our case we have also a measure prefactor but it is positive); the proof can be found in Ref.\ \cite{SZ}. 

For $|\tilde x-\tilde x'|\gg 1$, the asymptotic limit of $u_\b$ is given by the $n=1$ term,
\ba
u_\b(x-x',\s) &\ \stackrel{|x-x'|^2\gg \s^{\b/\g}}{\sim} \  &\left[\frac{\sin(\pi\g)}{\pi}\frac{\Gamma(2\g+1)}{\Gamma(\b+1)}\right]\nonumber\\
&&\times\frac{\s^{\b}}{|x-x'|^{2\g+1}}\,.\label{xgg}
\ea
This expression implies that all moments $\langle |x-x'|^a\rangle$ are finite if $0<a< 2\g<2$.

\begin{figure}
\centering
\includegraphics[width=8cm]{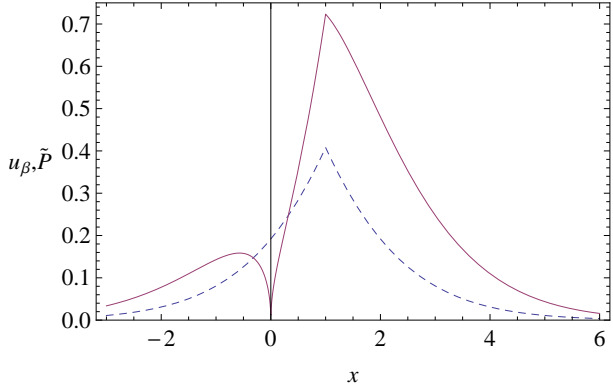}
\caption{\label{fig2} The probability densities \Eq{ub} (dashed curve) and $\tilde P=u_\b/v_\a(x)$ (solid curve) with $\b=1/2=\a$, $x'=1$ and $\s=1$. The plot of $P=u_\b/\sqrt{v_\a(x)v_\a(x')}$ is very similar to the one for $\tilde P$.}
\end{figure}

When $\g=1$, the above expressions are ill defined and one should find the solution in a different way. Using the Laplace-transform method, one can show \cite[Corollary 6.5]{KST} that, if $0<\b<1$, the solution is
\be\label{ub}
u_\b(x-x',\s) = \frac{\s^{-\frac{\b}{2}}}{2}\, W\left(-\tfrac{\b}{2};1-\tfrac{\b}{2};-|\tilde x-\tilde x'|\right)\,,
\ee
where
\be
W(a;b;z) := \sum_{n=0}^{+\infty} \frac{z^n}{n! \Gamma(a n+b)}
\ee
is Wright's function. An exact solution also exists for $1<\b<2$, which coincides with the one above if $\p_\s u_\b(x-x',\s)|_{\s=0}=0$ is imposed \cite[Corollary 6.6]{KST}. The solution \Eq{ub} and its fractional-space counterpart are plotted in Fig.\ \ref{fig2}.


\subsection{L\'evy process ($\b=1$, $0<\g<1$, $\cS=0$)}
\label{ssect:3.3}

The diffusion equation associated with L\'evy processes is integer in diffusion time and fractional in the spatial generator (i.e., the Laplacian). Our Laplacian is not the Riesz operator as in standard L\'evy processes (except in $D=1$) but the physics is qualitatively the same. The diffusion equation is then
\be\label{levy}
\left(\p_\s-\cK_{\g,\a}^{\rm E}\right)u_1=0\,.
\ee
When $\b=1$, Eq.~\Eq{kkk} is the product of $D$ L\'evy distributions $\rme^{-|k|^{2\g}}$ with L\'evy index $2\g$ \cite[Sec.\ 4]{Zas3}. The expression of the heat kernel is ($D$ copies of) Eq.~\Eq{hek1} with $\b=1$:
\ba
u_1(x-x',\s) &=& \prod_{\mu=1}^D\left\{\sum_{n=0}^{+\infty}\left[(-1)^{n+1}\frac{\sin(\pi\g n)}{\pi}\frac{\Gamma(2\g n+1)}{\Gamma(n+1)}\right]\right.\nonumber\\
&& \left.\times\frac{\s^{n}}{|x_\mu-x_\mu'|^{2\g n+1}}\right\}\,.\label{hekD}
\ea
To get an asymptotic expression for $x\sim x'$, we notice \cite{WeZ2} that the integrand in the momentum expression for $u_1$ is even and one can take the real part of $\rme^{\rmi k(x'-x)}$. Expanding the cosine, we obtain in one dimension
\ba
&&\int_{-\infty}^{+\infty}\frac{\rmd k}{2\pi}\,\rme^{-\s|k|^{2\g}}\,\rme^{\rmi k(x'-x)}\nonumber\\
&&\quad =
\sum_{m=0}^{+\infty} \frac{(-1)^m}{\pi(2m)!}\,(x-x')^{2m} \int_{0}^{+\infty}\rmd k\,\rme^{-\s|k|^{2\g}}k^{2n}\nonumber\\
 &&\quad =
\sum_{m=0}^{+\infty} \frac{(-1)^m}{(2m)!}\,\frac{\Gamma\left(\frac{2m+1}{2\g}\right)}{2\pi\g} \,\s^{-\frac{2n+1}{2\g}}\,(x-x')^{2m}\,,\nonumber
\ea
where we used \cite[Eq.~3.478.1]{GR}. Then,
\ba
u_1(x,x',\s) &=& \prod_{\mu=1}^D u_1(x_\mu-x_\mu',\s)\nonumber\\
 &=&\s^{-\frac{D}{2\g}}\prod_{\mu=1}^D\left\{\sum_{m=0}^{+\infty} \left[\frac{(-1)^m}{(2m)!}\, \frac{\Gamma\left(\frac{2m+1}{2\g}\right)}{2\pi\g}\right]\right.\nonumber\\
 &&\left.\vphantom{\prod_{\mu=1}^D}\times\s^{-\frac{m}{\g}}\,(x_\mu-x_\mu')^{2m}\right\}\,.\label{sol1}
\ea
Equation \Eq{levy} can be modified to the case where the spatial generator is the Riesz--Bessel operator \cite{ARAG}.


\subsection{Fractional-time diffusion equation ($0<\b<1$, $\g=1$, $\cS=0$)}\label{ftde}

Some subdiffusive processes are characterized by a transport equation with fractional diffusion operator and a second-order ordinary Laplacian \cite{Zas3,MLP,Nig86,Wis86,ScW,Ma95a,Mai96},\footnote{Often it is also described by another transport equation, called a bi-fractional or fractional Fick equation, where the diffusion equation is ``redistributed,'' $(\p_\s-\p_\s^{1-\b}\p_x^2)P=0$ \cite{MeK,MCK}. For Caputo derivatives these two formulations coincide, while for the Riemann--Liouville derivative a source term must be added to the first equation.}
\be\label{fbm}
\left(\p_\s^\b-\N_x^2\right)u_\b=0\,.
\ee
A nontrivial measure dependence will be present in the fractional-space case, but this generalization will be immediate. When $\g=1$, the integral in Eq.~\Eq{diffkernel2} can be factorized into a radial and an angular part. Eventually, $u_\b$ can be cast as a Fox function $H_{2,2}^{2,0}$ for $0<\b<1$ \cite[Corollary 6.7]{KST}, or as a Meijer $G$-function as we shall do later; if one assumes that $\p_\s u_\b(x-x',\s)|_{\s=0}=0$, the same solution holds also for $1<\b<2$ \cite[Corollary 6.8]{KST}. 
$u_\b$ (hence, $P$) is nonnegative definite for all these values of $\b$, including $\b=2$ \cite{MLP,BO1}.

\begin{figure}
\centering
\includegraphics[width=8cm]{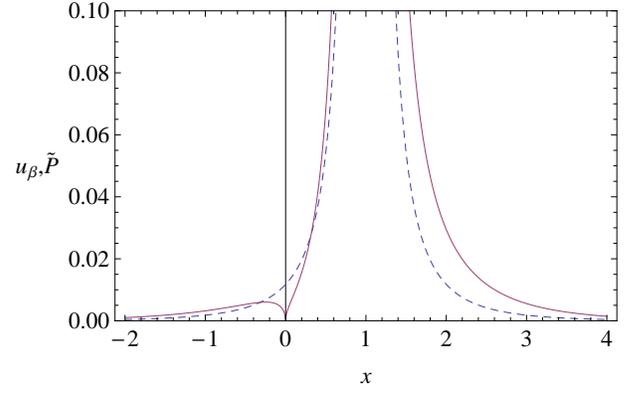}
\caption{\label{fig3} The solution \Eq{u12ok} of the quartic diffusion equation \Eq{mitF}  ($\b=1/2$, dashed curve) and $\tilde P=u_{1/2}/v_\a(x)$ (solid curve) with $D=4$, $\a=1/2$, $x'=1$ and $\s=1$. $\tilde P$ and $u_{1/2}$ are non-negative definite and can be interpreted as probability distributions.}
\end{figure}

Another formulation, more convenient to study the analytic properties of $P$, is the following \cite{OB2}. In one dimension, if $0<\b\leq 1$, the solution of
\bs\label{deb}\ba
&&(\p_\s^\b-\p_x^2)u_\b(x-x',\s)=0\,,\\
&&u_\b(x-x',0)=\delta(x-x')\,,\qquad 0<\b\leq 1,
\ea\es
is 
\be
u_\b(x-x',\s)=\int_0^{+\infty}\rmd s\,\frac{\rme^{-\frac{s^2}{4\s}}}{\sqrt{\pi\s}}\,u_{2\b}(x-x',\s)\,,
\ee
where $s$ is a parameter with engineering dimension $[s]=[\s]/2=-1/\b$ and $u_{2\b}$ is the solution of the analogous problem
\be\label{la0}
(\p_\s^{2\b}-\p_x^2)\,u_{2\b}(x-x',\s)=0\,,
\ee
with initial condition
\be\label{ic01}
u_{2\b}(x-x',0)=\delta(x-x')\,,\qquad 0<\b\leq \tfrac12
\ee
or
\bs\label{ic02}\ba
u_{2\b}(x-x',0)&=&\delta(x-x')\,,\\
\p_\s u_{2\b}(x-x',\s)\big|_{\s=0}&=&0\,,\qquad \tfrac12<\b\leq 1\,.
\ea\es
Clearly, the initial condition in Eq.\ \Eq{deb} is respected. In particular, when $\b=1/2$ the solution is the integral product of two Gaussians,
\be\label{twoG}
u_\frac{1}{2}(x-x',\s)=\int_0^{+\infty}\rmd s\,\frac{\rme^{-\frac{s^2}{4\s}}}{\sqrt{\pi\s}}\frac{\rme^{-\frac{(x-x')^2}{4s}}}{\sqrt{4\pi s}}\,,
\ee
and the generalization to $D$ dimensions is straightforward. The distribution
\be\label{u12}
u_\frac{1}{2}(r,\s)=\int_0^{+\infty}\rmd s\,\frac{\rme^{-\frac{s^2}{4\s}}}{\sqrt{\pi\s}}\frac{\rme^{-\frac{r^2}{4s}}}{(4\pi s)^{\frac{D}{2}}}
\ee
is the solution of the diffusion equation
\bs\label{de12}\ba
&&\left(\p_\s^\frac{1}{2}-\N_x^2\right)u_\frac{1}{2}(x-x',\s)=0\,,\\
&& u_\frac{1}{2}(x-x',0)=\delta(x-x')\,,\qquad 0<\b\leq 1\,.
\ea\es
All the above solutions are manifestly non-negative definite and normalized, 
\be
u_\b\geq 0\,,\qquad \int_{-\infty}^{+\infty}\rmd^Dx\,u_\b(x-x',\s)=1\,,
\ee
so they can be interpreted as probability densities. Under the rescaling to dimensionless variables $\tilde x = \s^{-\b/2}x$, $\tilde s = \s^{-\b/2}s$, the $\s$ dependence of $u_\b$ can be factorized as
\be
u_\b\propto \s^{-\frac{D\b}{2}}\,.
\ee
The integral in Eq.\ \Eq{u12} can be done exactly to yield a Meijer $G$ function:
\be\label{u12ok}
u_\frac{1}{2}(r,\s)= \frac{1}{4\pi^3 \sqrt{s}\, r^2}\,G_{00}^{30}\left(\left.\frac{r^4}{256\s}~\right|~\begin{matrix} 0 & 0 & 0\\ 0 & \frac12 & 1\end{matrix}\right)\,,
\ee
which is positive definite (Fig.\ \ref{fig3}).

When $\g=1$, Eq.~\Eq{xgg} vanishes and all moments with $a>0$ are finite. An explicit determination of the moments for even $a$ \cite{SZ} makes use of the diffusion equation. In fact,
\ba
\p^\b_\s \langle |x-x'|^a\rangle_\mu &\stackrel{\Eq{mome}}{=}&  \int_{-\infty}^{+\infty}\rmd^D x\,|x_\mu-x_\mu'|^a\,\p^\b_\s u_\frac{1}{2}(x-x',\s)\nonumber\\
&=&  \int_{-\infty}^{+\infty}\rmd^D x\,|x_\mu-x_\mu'|^a\,\cK_{\g,1}^{\rm E} u_\frac{1}{2}(x-x',\s)\nonumber\\
&=&  \int_{-\infty}^{+\infty}\rmd^D x\,u_\frac{1}{2}(x-x',\s)\,\cK_{\g,1}^{\rm E}|x_\mu-x_\mu'|^a,\nonumber
\ea
so that for $\g=1$
\ba
\p^\b_\s \langle |x-x'|^a\rangle_\mu &=& \int_{-\infty}^{+\infty}\rmd^D x\,u_\frac{1}{2}(x-x',\s)\,\p_\mu^2|x_\mu-x_\mu'|^a\nonumber\\
&=&a(a-1)\int_{-\infty}^{+\infty}\rmd^D x\,u_\frac{1}{2}(x-x',\s)\nonumber\\
&&\qquad\qquad\times|x_\mu-x_\mu'|^{a-2}\nonumber\\
&=& a(a-1)\langle |x-x'|^{a-2}\rangle_\mu\,.
\ea
Using the fact that (e.g., \cite{frc1})
\be
\s^{b-\b}=\frac{\Gamma(b-\b+1)}{\Gamma(b+1)}\p^\b_\s \s^b\,,\qquad b\neq 0\,,\label{speci}
\ee
we obtain the second moment ($a=2$)
\be
\langle |x-x'|^2\rangle_\mu = \frac{2}{\Gamma(\b+1)}\,\s^\b
\ee
and, by recursion, all even moments:
\be
\langle |x-x'|^{2n}\rangle_\mu = \frac{\Gamma(2 n+1)}{\Gamma(\b n+1)}\,\s^{\b n}\,,\qquad n\in\mathbb{N}\,.
\ee
Odd moments vanish, $\langle |x-x'|^{2n+1}\rangle_\mu=0$. In particular, the average of the square Euclidean distance $r^2$ covered by the particle is a function of diffusion time:\footnote{In fractional spaces, the natural distance is a $2\a$-norm \cite{frc1} but here, for simplicity, we use a $(2\a p)$-norm with $p=1/\a\geq 1$, which is topologically equivalent. Restoring measure factors and extending the discussion to fractional spaces does not entail further difficulties.}
\be\label{r2}
\langle (r-\langle r\rangle)^2\rangle =\frac{2D}{\Gamma(\b+1)}\,\s^\b\,.
\ee


\subsection{Iterated Brownian motion ($\b=1$, $\g=2$, $\cS\neq 0$)}

An example of particular importance for us is \emph{iterated Brownian motion} (IBM) \cite{OB2,Fun79,DeM,Bur92,Bur94,KL1,KL2,CCFR,KL3,AZ,All02,DeB04,OB1,Nan05,BMN1,BO3,Nan08,BMN2,BOS}. To illustrate iterated Brownian motion, we set $\a=1$ and consider the ordinary $D$-dimensional higher-order operator $\N^n$, Eq.~\Eq{laplan}. Given a process $X(\s)$, let $E$ be the expectation associated with $X(0)=x'$. Then, the function $u(x-x',\s)=E\{f[X(\s)]\}$ solves a certain diffusion equation with initial condition $u(x-x',0)$, where $x'$ is fixed. For an ordinary Brownian motion $X(\s)=B(\s)$, the distribution $u$ obeys the diffusion equation
\be\label{de1}
\left(\p_\s-\N^2_x\right)u(x-x',\s)=0\,,\qquad u(x-x',0)=\delta(x-x')\,.
\ee
Let $B_\pm$ be two independent standard Brownian motions defining the two-sided Brownian process
\be
X(t):=\left\{ \begin{matrix} B_+(t)\,,&\quad t\geq 0\,,\\
                             B_-(-t)\,,&\quad t< 0\,,\end{matrix}\right.
\ee
where $t\in\mathbb{R}$. Let $B$ be another independent Brownian motion. The IBM is the process defined as
\be
X_{\rm IBM}(\s):=X[B(\s)]\,,\qquad \s\geq 0\,,
\ee
where the Brownian motion $B$ acts as a clock to the two-sided motion. For this reason, iterated Brownian motion is often called also Brownian-time Brownian motion. Another definition of IBM is, given two independent Brownian processes $B_{1,2}$,
\be\label{biter2}
X_{\rm IBM}'(\s):=B_1[|B_2(\s)|]\,,\qquad \s\geq 0\,.
\ee
One can extend these definitions to multiple iterations of $n$ processes. If $X=X_{\rm IBM}$ or $X=X_{\rm IBM}'$, the diffusion equation is \cite{AZ,All02,DeB04,BMN1,OZ}
\be\label{mitF}
\left(\p_\s-\N^4_x\right)u(x-x',\s)=\frac{1}{\sqrt{\pi\s}}\N^2_x u(x-x',0)
\ee
in $D$ dimensions. Since the process is non-Markovian, the initial condition $u(x-x',0)$ explicitly acts as a source. Other processes than Brownian motions can be composed together and generate different diffusion equations (see \cite{BOS} for an account). In Sec.\ \ref{tele} we shall consider telegraph processes \cite{BO1,OB2,OB1,OZ,HO}.

A remarkable duality shows that fractional and iterated Brownian motions can be identified. Namely, Eq.~\Eq{u12} is a solution of \emph{both} \Eq{de12} \emph{and} \Eq{mitF}. In general, there exists a triple connection between fractional diffusion equations with fractional time $\s$, higher-order diffusion equations with integer time, and iterated processes \cite{AZ,DeB04,OB1,BMN1,Nan08,BOS}.

When the kinetic term is a second-order differential operator, the generator of the IBM is a ``half derivative'' which can be rigorously defined \cite{AZ}. Intuitively, this stems from the iteration of the fractional diffusion equation \Eq{de12}. Let $u_{1/2}$ be the solution of Eq.~\Eq{de12}. Applying $\p_\s^{1/2}$ twice and using the property $\p^{1/2}\p^{1/2}=\p$, valid for Caputo derivatives [see, e.g., Eqs.\ (2.50)--(2.53) of Ref.\ \cite{frc1}], we have for $x-x'\neq 0$
\ba
\p_\s u_{\frac12} &=& \p_\s^{\frac12}\p_\s^{\frac12} u_{\frac12}\nonumber\\
           &\ \stackrel{\Eq{de12}}{=}\ & \p_\s^{\frac12}\N^2_x u_{\frac12}\nonumber\\
           &\ \stackrel{x-x'\neq 0}{=}\ & \N^2_x\p_\s^{\frac12} u_{\frac12}\nonumber\\
           &=& \N^4_x u_{\frac12}\,.
\ea
For $x\sim x'$, the third step is not correct and one can check that $u_{1/2}$ is indeed a solution of \Eq{mitF} in a weak sense (i.e., by integrating the diffusion equation with a test function); this is the point where the source arises. The interested reader can find the proof in the references cited above.

The equivalence between fractional and iterated Brownian motion can play a role in the interpretation of quantum geometry at the UV fixed point, including QEG \cite{CES}. For this reason, it is important to stress the physical meaning of transport equations such as \Eq{de12} and \Eq{mitF}. There do exist physical systems which are associated with these hyperbolic equations. In fact, iterated Brownian motion provides one stochastic description (among others) of diffusion in cracks \cite{KL3,BuK}. Intuitively, it consists in the Brownian diffusion of a particle trapped in a random fractal set (a ``crack'') whose pattern resembles the graph of a Brownian motion. In the interpretation of Eq.~\Eq{biter2} in the crack model, large increments of $B_1$ over short intervals correspond to small-width spots along the crack, which appear with regularity.

More generally, time fractional derivatives are involved in diffusion equations describing transport on fractals \cite{MeK,MGN,GiR,MeN}. Both diffusion and the medium are then irregular.


\subsection{Other higher-order diffusion equations}\label{s5}

Without the source term, Eq.~\Eq{mitF} reduces to the prototypical higher-order hyperbolic equation
\be\label{x4}
(\p_\s-\N_x^4)u=0\,.
\ee
In contrast to hyperbolic diffusion equations \cite{OZ,HO}, parabolic equations such as
\be\label{+x4}
(\p_\s+\N_x^4)u=0
\ee
describe unconventional diffusion processes.\footnote{Parabolic equations of the type \Eq{+x4} were studied in Refs.\ \cite{OB2,BOS,OZ,HO,Kri60,Hoc79,BO2} and generalized to higher-order operators $\N_x^n$ and several other forms.} At first, such a differential equation seems unphysical: changing the relative sign between spatial and time derivatives is tantamount to asking the ink in a bottle of water to condense back to a single drop. The solution $u$ is not positive definite and cannot be interpreted as a probability density. These equations, anyway, come from composite processes just as their hyperbolic counterparts. In particular, they describe diffusion associated with the original formulation of the IBM \cite{Fun79,HO} (but not with IBM as later formulated and presented above) or, in the case of odd-order kinetic operators $\N_x^{2n+1}$, with the composition of Brownian motion with stable processes \cite{HO}.


\subsection{Spectral and walk dimension}\label{f5}

Let us consider fractional spacetimes with measure $v_\a$. The return probability is defined as the spatial average of $P$, i.e., the trace of the heat kernel per unit volume,
\ba
\cP(\s) &:=& \frac{1}{\cV_\a}\int_{-\infty}^{+\infty}\rmd\vr_\a(x)\,P(x,x,\s)\nonumber\\
&=&\frac{1}{\cV_\a}\int_{-\infty}^{+\infty}\rmd\vr_\a(x)\int_{-\infty}^{+\infty}\rmd\tau_{\a'}(k)\, f_k(\s)\, |\bE(k,x)|^2\nonumber\\
&=&\frac{1}{(2\pi)^D}\frac{1}{\cV_\a}\int_{-\infty}^{+\infty}\rmd^Dx\int_{-\infty}^{+\infty}\rmd^Dk\, f_k(\s)\,,\label{retprf}
\ea
where $\cV_\a:=\int\rmd\vr_\a(x)$ is a divergent total volume prefactor. Under the rescaling \Eq{reskx}, we get
\be\label{ksa}
\cP(\s) = A\,\s^{-\frac{D\a\b}{2\g}}\,,
\ee
where $A$ is a numerical constant:
\be\label{A}
A=\cI_{\b,\g}\frac{\int_{-\infty}^{+\infty}\rmd^D\tilde x}{\int_{-\infty}^{+\infty}\rmd^D\tilde x\,v_\a(\tilde x)}\,,\quad \cI_{\b,\g}:= \int_{-\infty}^{+\infty}\frac{\rmd^D\tilde k}{(2\pi)^D}\, f_{\tilde k}(1)\,.
\ee
The $\s$ dependence of $\cP$ comes exclusively from (i) the rescaling \Eq{reskx} and (ii) the volume prefactor in the denominator. In particular, it depends on the topological dimension $D$ of position and momentum space. It does not depend on (i$'$) the measure weight $w(k)$ in momentum space, nor on (ii$'$) the relative sign between diffusion and kinetic operators (hyperbolic or parabolic diffusion equation), nor on (iii$'$) the presence of friction or source terms, as one can convince oneself by a direct inspection (e.g., \cite{OB2,BOS} and references therein). Consequently, all these properties will be inherited by the spectral dimension.

The dimensionless coefficient in Eq.~\Eq{ksa}, formally indeterminate, can be regularized so that $A=1$. For simplicity, we specialize to the isotropic case $\a_\mu=\a$. The $\tilde x$-dependent integrals must be regularized since they diverge at $\pm\infty$. Defining two parameters $\e=\e(\ve)$ and $0<\ve\ll 1$, we write
\ba
\frac{\int_{-\infty}^{+\infty}\rmd^D\tilde x}{\int_{-\infty}^{+\infty}\rmd^D\tilde x\,v_\a(\tilde x)} &=& \left[\Gamma(\a)\frac{\int_0^{+\infty}\rmd\tilde x}{\int_0^{+\infty}\rmd\tilde x\,\tilde x^{\a-1}}\right]^D\nonumber\\
&:=& \lim_{\ve\to 0^+} \left[\Gamma(\a)\frac{\int_0^{\frac1\e}\rmd\tilde x}{\int_0^{\frac{1}{\ve(\e)}}\rmd\tilde x\,\tilde x^{\a-1}}\right]^D\nonumber\\
&=& \lim_{\ve\to 0^+}\left[\Gamma(\a+1)\frac{\ve^\a(\e)}{\e}\right]^D\,.\nonumber
\ea
Assuming that $\cI_{\b,\g}>0$, we can take
\be\label{regupro}
\e=\Gamma(\a+1)\cI_{\b,\g}^{1/D}\ve^\a\,.
\ee
The arbitrariness of this procedure does not affect the physics, which is encoded in the spectral dimension
\be\label{spedi}
\ds := -2\frac{\rmd\ln \cP(\s)}{\rmd\ln\s}\,,
\ee
leading to the final result
\be\label{dsbg}
{\ds=\frac{\b}{\g}\,\dh\,.}
\ee
We can distinguish three cases:

(1) For $\b=\g$ (including normal diffusion, $\b=\g=1$), $\ds=\dh$, whether it be realized by integer or fractional differential operators $\p_\s^\b$ and $\cK_{\b,\a}$. The important point is that the order of the diffusion operator is the natural one, i.e., half that of the Laplacian. $\cM_v^D$ is a fractal.

(2) When $\b<\g$, diffusion is anomalous and the spacetime $\cM_v^D$ can again be regarded as a fractal. This happens, in particular, for integer-order Laplacian and fractional diffusion ($\b<1$ is assumed).

(3) As remarked in Ref.\ \cite{frc1}, the superdiffusion case $1\geq \b>\g$ does not correspond to a fractal, since $\ds>\dh$. In particular, for integer diffusion ($\b=1$) this is a L\'evy process. When $\b=1$ and $\g=\a$, the spectral dimension coincides with the topological dimension of space, $\ds=D$.

In the case $\g=\a$, the results are summarized in Table \ref{tab1} for $\ds$ and the walk dimension 
\be\label{dw}
{\dw:=2\frac{\dh}{\ds}=2\frac{\g}{\b}\,.}
\ee
The walk dimension is also defined through the scaling of the second moment, $\langle r^2\rangle\sim \s^{2/\dw}$. By a purely dimensional argument, one can infer that $2/\dw=\b/\g$, in agreement with Eqs.~\Eq{r2} and \Eq{dw}. When $\dw>2$, $\dw=2$, $\dw<2$, and $\dw=1$ the process is, respectively, subdiffusive ($\g>\b$), normal ($\g=\b=1$), superdiffusive ($\g<\b$), and ballistic ($2\g=\b$).
\begin{table}[ht]
\caption{\label{tab1}Spectral dimension $\ds$ and walk dimension $\dw$ of spacetime $\cM_v^D$ for different harmonic structures (Laplacians and diffusion equations). $\cM_v^D$ is fractal only if $\dw\geq 2$.}
\centering
\begin{tabular}{c||cc||cc}\hline
                  & \multicolumn{4}{|c}{Laplacian}  \\ \hline
       			      & \multicolumn{2}{|c||}{$\cK=\cK_{1,\a}$}  & \multicolumn{2}{|c}{$\cK_{\a,\a}$} \\\hline\hline
$\p_\s$           & $\ds=\dh$     &    $\dw=2$    & $\ds=D\geq\dh$           &        $\dw=\a<2$  \\ 
$\p_\s^\b$        & $\ds=\b\dh\leq\dh$ & $\dw=\frac{2}{\b}>2$ &   $\ds=\frac{\b}{\a}\dh$ & $\dw=2\frac{\a}{\b}$ \\\hline
\end{tabular}
\end{table}

A final caveat should be stressed. In the regularization procedure, the choice \Eq{regupro} was possible under the assumtpion that $\cI_{\a,\g}>0$. Otherwise, not only the limit of the regulator but also the return probability could be ill defined, taking negative or complex values. In turn, the positive definiteness of the probability depends on the Hermiticity of the Laplacian. We have already seen that there exists a parameter range such that $\cI_{\a,\g}>0$ for the ``Riesz choice'' of the coefficients in Eq.~\Eq{cka2}.


\section{Multiscale processes and spacetimes}\label{f6}

In quantum-gravity theories, the spectral dimension is found to depend on the physical length scale $\ell$ one is probing. In the previous section we found the spectral dimension of a fractal model with no scale dependence, so that $\ds$ is constant. One needs to generalize the discussion to a multiscale geometry. The latter will be multifractal only in certain ranges of parameter space, where $\ds\leq\dh$. Some basic properties of multiscale measures and systems \cite{GrP,JKLOS,HJKPS,PV,Har01,Fal03} were reviewed in Ref.\ \cite{frc2}.

Before dealing with multifractional spacetimes, we clarify the problem of multiscaling at the level of the diffusion equation, with $\a=1$ and $D=1$. The generalization of Eq.\ \Eq{1dde} to a multiscale process (and a source) is simply achieved by summing over all possible values of $\b$ or $\g$: 
\be\label{muf1}
\left(\sum_\b \xi_\b\p_\s^\b-\sum_\g \zeta_\g\p^{2\g}_{|x|}\right)\,u(x-x',\s)=\cS(x-x',\s)\,,
\ee
where $\xi_\b$ and $\zeta_\g$ are dimensionful couplings which depend on the characteristic scales of the system. The parameters $\b$ and $\g$ can be also let vary continuously, thus having integrations instead of sums \cite{Zas3,Zas00}, or both \cite{Zas1}. In particular, by making use of the so-called distributed-order fractional derivatives \cite{Cap69,Cap95,BT1,BT2,CGS,LoH02,Koc1,Koc2} one can construct and analyze multiscale fractional diffusion equations \cite{CGS,Koc1,Cap01,CKS,Nab04,MPG,CGGKS,CSK}. Here we limit the discussion to a simple case where the sums in Eq.\ \Eq{muf1} are replaced by an integral over a length parameter $\ell$,
\be\label{muf2}
\int\rmd\ell\left[\xi(\ell)\p_\s^{\b(\ell)}-\zeta(\ell)\p^{2\g(\ell)}_{|x|}\right]\,u(x-x',\s,\ell)=\cS(x-x',\s)\,,
\ee
where the length $\ell$ is \emph{not} identified with (a power of) the diffusion parameter $\s$ and the range of $\ell$ may be chosen appropriately (for instance, $\ell \in[\ell_0,+\infty)$ in the presence of a cut-off). To get the solution of Eq.\ \Eq{muf2} for $\cS=0$, it is sufficient to solve for the integrand,
\be\label{muf3}
\left[\p_\s^{\b(\ell)}-\zeta(\ell)\p^{2\g(\ell)}_{|x|}\right]\,u(x-x',\s,\ell)=0\,,
\ee
where $\xi(\ell)$ has been absorbed into $\zeta(\ell)$. From Eq.\ \Eq{dsbg}, the spectral dimension would be ($\dh=1$ here)
\be\label{dsbgell}
\ds(\ell)=\frac{\b(\ell)}{\g(\ell)}\,.
\ee
The sum representation is more convenient unless one has an argument to choose the profiles $\b(\ell)$ and $\g(\ell)$ in a specific way (as we shall do in Sec.\ \ref{mfsts}). Also, typically there is only a finite number of terms in physical systems \cite{HJKPS,PV}. When the sums are finite, Eq.~\Eq{muf1} does admit analytic solutions in several cases.

It is important to stress that \emph{the number of scales in the system determines the number of plateaux or asymptotic regimes in the profile of the spectral dimension} $\ds$. This number, corresponding to the number of H\"older exponents in a self-similar measure, is given by
\ba
&&\text{(no.\ of asymptotic regimes of $\ds$)}\nonumber\\
&&\quad= \text{(no.\ of $\b$'s)}+\text{(no.\ of $\g$'s)}-1\nonumber\\
&&\quad= \text{(no.\ of characteristic scales of the system)}+1\,.\nonumber\\
\ea
Take, for instance, the multifractional-time diffusion equation with $N$ diffusion operators and the ordinary Laplacian ($\g=1$). Then, the number of characteristic regimes of the multifractal spectrum is just the number of $\beta$'s. Assume that $\s$ has dimension of a squared length, so that the couplings $\xi_\b$ can be written in terms of $N$ length scales $\ell_n$, $n=1,\dots,N$:
\be
\xi_{\b_n}=\ell_n^{2\b_n-2}\,.
\ee
If the $N$th operator has $\b=1$, the diffusion process is governed by
\be
\left[\p_{\bar\s}+\sum_{n=1}^{N-1}\left(\frac{\ell_n}{\ell_N}\right)^{2(\b_n-1)}\p^{\b_n}_{\bar\s}-\ell_N^2\p^2_x\right]u=0\,,
\ee
where $\bar\s:=\s/\ell_N^2$ is dimensionless. Later we shall better specify the scale dependence of the coefficients.

The discrepancy between the number of fundamental scales and the number of regimes is due to the fact that a multiscale phenomenon is always defined by the relative size of the scales, i.e., by a hierarchy. This means that we can choose any of the $N$ scales $\ell_n$ to represent the variable scale $\ell$ probed by a measurement. Thus, there are $N-1$ (not $N$) scales with the physical meaning of characteristic lengths. The spectral dimension is fixed when $N=1$; for $N=2$ it has two asymptotic values $\ds\sim {\ds}_{1,2}$ (with a monotonic\footnote{A local extremum, for instance a minimum value ${\ds}_{,{\rm min}}$ at some point $0<\ell_{\rm min}<+\infty$ would signal the presence of another scale because this feature could not be removed by a finite conformal rescaling of diffusion time.} transient phase in between) in the regimes $\ell_2\ll\ell_1$ and $\ell_2\gg\ell_1$; for $N=3$ there will be two fundamental scales and a plateau $\ds\sim {\ds}_3$ at intermediate scales; and so on.

We need only one scale to obtain a multifractal or, more generally, a geometry with scale-dependent spectral dimension. Therefore, as a first approximation, it is sufficient to study diffusion equations with two diffusion operators $\p_\s^{\b_1}$ and $\p_\s^{\b_2}$,\footnote{Some remarks on the case with a finite but arbitrary number $N$ of diffusion operators are given in Ref.\ \cite{AAMR}.} or with two Laplacians $\p^{2\g_1}_{|x|}$ and $\p^{2\g_2}_{|x|}$. Certain realizations of QEG can be regarded as a two-scale system, where there is an intermediate genuine plateau  \cite{CES}.


\subsection{Multiscale L\'evy process}

The first example is an interaction of Gaussian and anomalous dynamics which can describe certain
turbulent media \cite{WeZ1,CGS}. The diffusion equation is
\bs\label{mufg}\ba
&&\left(\p_\s-\p_x^2-\zeta\p^{2\g}_{|x|}\right)\,u(x-x',\s)=0\,,\\
&& u(x-x',0)=\de(x-x')\,,\qquad 0<\g<1\,,
\ea\es
where $[\s]=-2$ and $\zeta$ is a coupling constant which, by a dimensional argument, can be written in terms of a characteristic length scale $\ell_*$:
\be
\zeta= \ell_*^{-2(1-\g)}\,,\qquad [\zeta]=2(1-\g)\,.
\ee
The Fourier transform of the spatial part is
\be
(k^2+\zeta|k|^{2\g})\tilde u(k,\s)=\ell_*^{-2}[(\ell_*k)^2+|\ell_* k|^{2\g}]\tilde u(k,\s)\,,
\ee
stating that the transport is normal at small scales $k^{-1}\ll \ell_*$ and of L\'evy type at large scales $k^{-1}\gg \ell_*$. At scales $k^{-1}\sim \ell_*$ the two behaviors compete. The normalized analytic solution can be written in different ways \cite{WeZ1}. For us, the following is convenient:
\bs\ba
&& u(x-x',\s) = \sum_{n=0}^{+\infty}(-1)^n y_n(\s)\,|\bar x-\bar x'|^{2n},\\ 
&& y_n(\s)=\frac{1}{(2n)!\pi}\int_0^{+\infty}\rmd \bar k\,\bar k^{2n}\rme^{-\frac{\s}{\ell_*^2}(\bar k^2+\bar k^{2\g})},
\ea\es
where all barred quantities are dimensionless:
\be
\bar x:= \frac{x}{\ell_*}\,,\qquad \bar k:=\ell_* k\,.
\ee
The coefficients $y_n$ can be expanded as a series for large and small $\ell_*^{-2}\s$. In particular, we are interested in the return probability and for $x=x'$ only the $n=0$ term contributes:
\be
\cP(\s)=y_0(\s)\sim\left\{\begin{matrix} \frac{1}{\sqrt{4\pi}} \left(\frac{\ell_*^2}{\s}\right)^{\frac12}\,,&\quad \s\ll\ell_*^2\\ \frac{1}{\sqrt{4\pi}} \left(\frac{\ell_*^2}{\s}\right)^{\frac{1}{2\g}}\,,&\quad \s\gg\ell_*^2\end{matrix}\right.\,.
\ee
Therefore, the spectral dimension is $\ds\sim 1$ at small scales and $\ds\sim 1/\g>1$ at large scales. This is consistent with Eq.~\Eq{dsbgell} with $\b(\ell)=1$ and a profile 
\be
\g(\ell)\sim\left\{\begin{matrix} 1\,,&\quad \ell\ll\ell_*\\
                                  \g<1\,,&\quad \ell\gg\ell_*\end{matrix}\right.\,.
\ee

From the perspective of quantum spacetimes, this model is multiscale but \emph{not} multifractal, since $\ds>\dh$. Profiles of $\ds$ over-shooting the Hausdorff and topological dimensions of space appear also in lattice-based \cite{COT} and noncommutative geometries \cite{AA}. Due to the presence of a characteristic scale (the noncommutative fundamental scale \cite{AA}, or the lattice cell size\footnote{In the case where the underlying lattice-type construction is taken as a regularization, these are lattice artifacts.} \cite{COT}, or the label-dependent length of the edges of a labeled graph \cite{COT}), the geometric information in the diffusion equation determines a nontrivial spatial generator. On the other hand, the diffusion operator is assumed to be the integer one $\p_\s$, so these models roughly mimic certain properties of L\'evy processes. In these cases, the flow is not monotonic since the scale acts as a minimum length cutoff, so $\ds\to 0$ in the UV. Since $\ds\to D$ at large scales, there is one local maximum in $\ds(\ell)$ (for labeled graphs there are, in fact, several extrema in $\ds$ \cite{COT}).


\subsection{Fractional telegraph process}\label{tele}

A fractional diffusion equation with multiple diffusion operators $\p^{\b_n}$ admits a neat stochastic interpretation in the case $N=2$. For the purpose, we recall some results on the so-called telegraph processes \cite{Gol51,Kac74,Ors85,Ors90}. After this short review, we shall make the connection with multiscale quantum spacetimes apparent.

A telegraph process $V$ in time $\s$ (here $[\s]=-1$) is defined as
\be
V(\s)=V(0)\,(-1)^{\cN(\s)}\,,
\ee
where $V(\s)$ is the velocity of a particle at time $\s$ running on the real line, $V(0)$ is the initial velocity which is $\pm c$ with equal probability, and $\cN$ is the cumulative number of events of a homogeneous Poisson process. Thus, the velocity of the particle flips direction at times obeying a Poisson distribution, hence the name ``telegraph.'' The position of the particle at time $\s$ is the integrated telegraph process
\be
T(\s)=V(0)\int_0^\s\rmd s\,(-1)^{\cN(s)}
\ee
and its probability distribution obeys the diffusion equation \cite{Gol51,Kac74}
\bs\label{tel1}\ba
&&(\p_\s^2+2\la\p_\s-c^2\p_x^2)u(x-x',\s)=0\,,\\
&& u(x-x',0)=\de(x-x')\,,
\ea\es
where $\la>0$ is the rate of the Poisson process. The variance of the process can be shown to scale as $\langle r^2\rangle\sim \s$ at large $\s$. When $\la,c\to+\infty$ and the ratio $\la/c^2$ remains constant, Eq.\ \Eq{tel1} reduces to the ordinary diffusion equation \cite{Kac74}. The limit $\la\to+\infty$ means that changes in the speed, abrupt in the telegraph process, take place continuously. 

The composition of a Brownian motion $B$ with an integrated telegraph process,
\be\nonumber
X_{{\rm DBM}}(\s):=\left\{ \begin{matrix} B[T(\s)]\,,&\quad T(\s)> 0\,,\\
                              \rmi B[-T(\s)]\,,&\quad T(\s)< 0\,,\end{matrix}\right.
\ee
is called delayed Brownian motion (DBM) \cite{HO}, and its probability distribution is governed by the quartic diffusion equation
\bs\label{tel2}\ba
&&\left(\p_\s^2+2\la\p_\s-\frac{c^2}{4}\p_x^4\right)u(x-x',\s)=0\,,\\
&& u(x-x',0)=\de(x-x')\,.
\ea\es
In the double limit $\la,c\to+\infty$, $\la/c^2\to {\rm const}$, the delayed Brownian motion reduces to Eq.\ \Eq{+x4}. Other combinations of telegraph and Brownian processes are possible, leading to different quartic diffusion equations \cite{OZ,HO}.

Finally, one can devise the fractional analog of the telegraph equation \cite{BO1,OB1}:
\be\label{tel3}
\left(\p_\s^{2\b}+2\la\p_\s^\b-c^2\p_x^2\right)u_{2\b}(x-x',\s)=0\,,
\ee
with initial condition given by Eq.\ \Eq{ic01} or \Eq{ic02}. 
 When $\la=0$, Eq.\ \Eq{tel3} reduces to the fractional diffusion equation \Eq{la0}, while for $\b=1$ it is the telegraph equation. Generalizations of Eq.\ \Eq{tel3} were inspected in Refs.\ \cite{OB2,SMH0}.

The analytic solutions of the telegraph equation \Eq{tel1}, of the delayed-Brownian-motion equation \Eq{tel2}, and of the fractional telegraph equation \Eq{tel3} were found, respectively, in Refs.\ \cite{Ors85,Ors90}, \cite{HO}, and \cite{OB1}. The latter, which is non-negative and unique, is given as a Fourier integral:
\ba
u_{2\b}(x-x',\s) &=& \frac{1}{2\pi}\int^{+\infty}_{-\infty}\rmd k\, \tilde u_{2\b}(k,\s)\,\rme^{-\rmi k x}\,,\label{fous}\\
\tilde u_{2\b}(k,\s) &=& \frac12\left[\frac{\eta_-}{\la+\eta_-}\,E_\b(\eta_+\s^\b)\right.\nonumber\\
&&\qquad\left.+\frac{\eta_+}{\la+\eta_+}\,E_\b(\eta_-\s^\b)\right]\,,
\ea
where $E$ is the Mittag-Leffler function \Eq{mil} and
\be
\eta_\pm = \pm\sqrt{\la^2-c^2k^2}-\la\,.
\ee
The variance of the process scales as $\langle r^2\rangle\sim \s^\b$ in the limit of large $\s$.

In the special case $\b=1/2$, the fractional telegraph equation becomes
\be\label{tel4}
\left(\p_\s+2\la\p_\s^{\frac12}-c^2\p_x^2\right)\,u_1(x-x',\s)=0\,.
\ee
Its solution can be written explicitly as a one-parameter integral:
\ba
u_1(x-x',\s) &=& \frac{1}{2c}\int_0^{+\infty}\rmd s\,\frac{\rme^{-\frac{s^2}{4\s}-\la s}}{\sqrt{\pi\s}}\left\{\vphantom{\frac12}\de(r-s)+\de(r+s)\right.\nonumber\\
&&\qquad\quad+\theta(s-r)\left[\la I_0\left(\la\sqrt{s^2-r^2}\right)\right.\nonumber\\
&&\qquad\left.\vphantom{\frac12}+\left.\p_s I_0\left(\la\sqrt{s^2-r^2}\right)\right]\right\}\,,\nonumber\\\label{BTsol}
\ea
where $r=|x-x'|/c$ and $I_0(z)=\sum_{n=0}^{+\infty} (z/2)^{2n}/(n!)^2$ is the modified Bessel function of the first kind and zeroth order.

In the double limit $\la,c\to+\infty$, $\la/c^2\to {\rm const}$, the process underlying \Eq{tel4} reduces to an iterated Brownian motion and, in fact, the limit of Eq.\ \Eq{BTsol} is \Eq{twoG}. The probability distribution $u_1$ is associated with a telegraph process with Brownian time,
\be
X_{{\rm FTP}}(\s)= T[|B(\s)|]\,,
\ee
which we call also \emph{fractional telegraph process}. The composite process $X_{\rm FTP}$ describes the random motion of a particle during a time interval of length $|B|$, so that at time $\s$ the particle is located in the random spatial interval $(-\s|B(\s)|,\s|B(\s)|)$ \cite{OB1}. This motion is governed by the multifractional diffusion equation \Eq{tel4}. Due also to the different orders of composition of $T$ and $B$, the fractional telegraph process $X_{{\rm FTP}}$ is different from the delayed Brownian motion $X_{{\rm DBM}}$.

We now end the digression into probability theory and recast these results in the language of multiscale spacetimes, as follows. We have to reshuffle the units of the variables and constants to agree with our past notation, where $[\s]=-2$. Setting $c=1$ and $\ell=1/(2\la)$ as the probed scale, Eq.\ \Eq{tel4} becomes
\be\label{tel5}
\left(\p_{\bar\s}+\frac{\ell_*}{\ell}\,\p_{\bar\s}^{\frac12}-\ell_*^2\p_x^2\right)u_1=0\,,
\ee
where, as before, $\bar\s=\ell_*^{-2}\s$ is dimensionless. In the limit $\ell\gg\ell_*$, diffusion in spacetime is Gaussian and described by a Brownian process. At small scales $\ell\ll\ell_*$, on the other hand, one reaches a regime where diffusion is fractional and described by an iterated Brownian motion. In between, diffusion in quantum spacetime obeys the law of a fractional telegraph process.

\begin{figure}
\centering
\includegraphics[width=8.6cm]{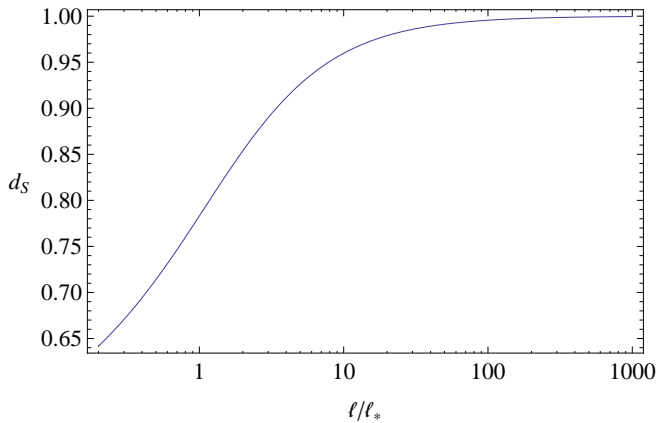}
\caption{\label{fig4} The spectral dimension $\ds(\ell)$ of a $D=1$ multifractal space whose stochastic properties are given by a fractional telegraph process.}
\end{figure}

The return probability
\ba
\cP(\s)&=& u_1(0,\s) = \frac{1}{\sqrt{\pi\s}}\left\{1+\frac{1}{4\ell}\int_0^{+\infty}\rmd s\,\rme^{-\frac{s^2}{4\s}-\frac{s}{2\ell}}\right.\nonumber\\
&&\qquad\qquad\quad\times\left.\vphantom{\int}\left[I_0\left(\frac{s}{2\ell}\right)+I_1\left(\frac{s}{2\ell}\right)\right]\right\}\label{BTsol0}
\ea
can be manipulated to give a sum of generalized hypergeometric functions with argument $\s/\ell^2=(\ell_*/\ell)^2\bar\s$. From that, we can obtain the profile of the spectral dimension $\ds(\ell)$, which is shown in Fig.\ \ref{fig4} (the portion of the curve below $\ds=0.65$ down to $\ds=1/2$ is not shown because of bad convergence of the code). As expected,
\be
\ds\sim\left\{ \begin{matrix} 1\,,&\quad \ell\gg\ell_*\,\qquad \mbox{(IR)}\,,\\
                              \frac{1}{2}\,,&\quad \ell\ll\ell_*\,\qquad \mbox{(UV)}\,.\end{matrix}\right.
\ee

The results above can be generalized to $D$ dimensions because the Fourier integral \Eq{fous} can be split into an angular and a radial integral, $\rmd^D k=\rmd\Omega_{D-1}\rmd k\,k^{D-1}$.  


\subsection{Multifractional measure}\label{mfsts}

All the above examples are realizations of multifractional spacetimes where $\a=1$ and the Hausdorff dimension of position space equals the topological dimension, $\dh=D$. The spectral dimension is anomalous either because of a fractional diffusion operator or due to a fractional Laplacian, or because of both. Inclusion of a nontrivial measure weight for position space makes $\dh$ anomalous, too. Formally, the extension of fractional to multifractional spacetimes is simply achieved by summing (or integrating) over all possible values of $\a$ \cite{fra4,frc2}. A multifractal action with a finite number of charges $\a$ thus reads
\be\label{muf}
S=\sum_{n=1}^N g_{n}\int\rmd^Dx\, v_{\a_n}(x)\,\cL_{\a_n,\g}\,,
\ee
where $g_n$ are some (dimensionful) couplings and $\cL_{\a,\g}$ is the Lagrangian density, possibly dependent on $\a$ and/or the order of the kinetic operator. The aim is to find the scale-dependent spectral dimension $\ds$. This entails a modification of the diffusion equation and the introduction of scale dependence in the parameters $\a$, $\b$, and $\g$. Due to its phenomenological character, we do not expect to obtain a unique multiscale extension. However, one can restrict the possibilities by an educated guess.

The general diffusion equation on the spaces described by \Eq{muf} is
\be\label{difef3}
\sum_{n=1}^N \left(\xi_n\p_\s^{\b_n}-\zeta_n\cK_{\g_n,\a_n}^{\rm E}\right)P(x,x',\s)=\cS(x,x',\s)\,,
\ee
given some initial condition $P(x,x',0)$. If we fix $\g_n=\g$ and order the $N-1$ scales of the system as $\ell_1<\ell_2<\dots<\ell_{N-1}$, we can argue that the coefficients $\zeta_n$ have the forms $\zeta_N=1$ and
\bs\label{zeta}\ba
&&\zeta_1(\ell)=\left(\frac{\ell_1}{\ell}\right)^{2\g}\,,\\
&&\zeta_n(\ell)=\left(\frac{\ell_n}{\ell-\ell_{n-1}}\right)^{2\g}\,,\qquad n=2,\dots,N-1\,,\nonumber\\
\ea\es
where $\ell=\ell_N>\ell_{N-1}$. To show this, we notice that the Laplacians all have the same order $2\g$, so the coefficients $\zeta_n$ all have the same scaling dimension. By fixing the scaling of the $\xi_n$ suitably, we can always make $\zeta_n$ dimensionless. This means, in particular, that we can write $\zeta_n$ as the ratio of some length scales, $\zeta_n=(l_{A,n}/l_{B,n})^q$. Without loss of generality, one can choose $q=2\g$ so that the spatial generator of the diffusion equation can be rendered dimensionless, in the form $\sum_n(l_{A,n})^{2\g}\cK_{\g,\a_n}$. Now, the $n$th term dominates over the others at scales $\ell\ll\ell_n$, so we could set $l_{A,n}=\ell_n$ and, tentatively, $l_{B,n}=\ell$. However, at scales smaller than $\ell_{n-1}$ the $(n-1)$th term takes the lead, so the smallest possible scale $\ell$ at which the $n$th term dominates is $\ell\sim\ell_{n-1}$. Therefore, the correct choice is $l_{B,n}=\ell-\ell_{n-1}$. In other words, the dimensional flow is always measured starting from the lowest of two scales $\ell_{n-1}$ to the next $\ell_n$, and relatively to the latter, which sets a gauge for the rods. Beyond the smallest scale $\ell_1$ there is nothing else to compare with and $\ell_0=0$. Since $\ell=\ell_N$ is the probed scales, $\zeta_N\equiv1$ by definition.

In the Gaussian case $\b=\g=1$ and just two entries ($N=2$, $\a_2=1$, $\a_1=\a_*$), dimensional flow is such that
\be\label{iruv}
\ds\sim\left\{ \begin{matrix} D\,,&\quad \ell\gg\ell_*\,\qquad \mbox{(IR)}\,,\\
                                 D\a_*=2\,,&\quad \ell\ll\ell_*\,\qquad \mbox{(UV)}\,,\end{matrix}\right. 
\ee
with no intermediate regimes in between. Explicitly, Eq.\ \Eq{difef3} becomes
\be\label{difef4}
\left[\p_\s^\b-\N_x^2-\left(\frac{\ell_*}{\ell}\right)^2\cK_{1,\a_*}^{\rm E}\right]P(x,x',\s)=\cS(x,x',\s)\,,
\ee
where we set $\zeta_2=1$ and $\zeta_1=\zeta=(\ell_*/\ell)^2$ according to Eq.\ \Eq{zeta}. When $\cS=0$, the UV and IR asymptotics of the solution are obvious.

Equations \Eq{difef3} and \Eq{difef4} are not much prone to manipulation, since fractional momentum transforms are not easily generalizable to the multifractional case when $D\geq 2$ \cite{frc3}. However, we can reinterpret the sum over $\a$ by imagining the dimensionless parameter $\a=\a(\ell)$ to depend on the probed scale. Equation \Eq{muf} can be recast as
\be
S=\int_0^{+\infty}\rmd\ell\,g(\ell)\int\rmd^Dx\, v_{\a(\ell)}(x)\,\cL_{\a(\ell),\g(\ell)}\,,
\ee
while the diffusion equation \Eq{difef3} generalizes Eq.\ \Eq{muf2} to
\ba
&&\int\rmd\ell\left[\xi(\ell)\p_\s^{\b(\ell)}-\zeta(\ell)\cK_{\g(\ell),\a(\ell)}^{\rm E}\right]P(x,x',\s,\ell)\nonumber\\
&&\qquad=\cS(x-x',\s).\label{muf4}
\ea
In the absence of source, it is sufficient to solve for the integrand,
\be\label{muf5}
\left[\p_\s^{\b(\ell)}-\zeta(\ell)\cK_{\g(\ell),\a(\ell)}^{\rm E}\right]\,P(x,x',\s,\ell)=0\,,
\ee
where, as before, we absorbed $\xi(\ell)$ in $\zeta(\ell)$. All the calculations of the case with fixed dimensionality are transposed with the replacement $(\a,\b,\g)\to (\a(\ell),\b(\ell),\g(\ell))$ and the spectral dimension is the generalization of Eq.\ \Eq{dsbgell}:
\be\label{dsbgellD}
\ds(\ell)=\frac{\b(\ell)}{\g(\ell)}\,\dh(\ell)\,,\qquad \dh(\ell)=D\a(\ell)\,.
\ee
We can plot the spectral dimension for any given profiles $\a(\ell),\b(\ell),\g(\ell)$. To obtain a multifractal, $\b/\g\leq 1$ throughout the whole evolution, so for simplicity we assume $\g=\b=1$. In Ref.\ \cite{frc2}, several \emph{Ans\"atze} were given for an $\a(\ell)$ with the desired asymptotic behavior \Eq{iruv}.\footnote{Equation (1.1) of Ref.\ \cite{frc2} has a typo and should read $\a(\ell)=1+(\a_*-1)/[1+(\ell_*/\ell)^{\a_*-1}]$.} They are all monotonic and lead to the same plot qualitatively, the only change being in the time-scale of the process. Here, however, we choose a profile which can be motivated as a realistic approximation of the sum \Eq{difef3} with $\g_n=1$ for all $n$. Consider first the $N=2$ case with $\a_1=\a_*$ and $\a_2=1$. In one dimension,
\ba
&&(\p_x^2+\zeta_1\cK_{1,\a_*}^{\rm E})P \nonumber\\
&&\quad= (1+\zeta_1)\left[\p_x^2-\left(1-\frac{1+\zeta_1\a_*}{1+\zeta_1}\right)\frac{1}{x}\p_x\right.\nonumber\\
&&\qquad\left.+\frac{\zeta_1}{1+\zeta_1}\frac{(1-\a_*)(3-\a_*)}{4x^2}\right]P\nonumber\\
&&\quad= \left[(1+\zeta_1)\cK_{1,\a_1(\ell)}^{\rm E}+\frac{\zeta_1}{1+\zeta_1}\frac{(1-\a_*)^2}{4x^2}\right]P,\label{useful}
\ea
where
\be\label{prof1}
\a_1(\ell):=\frac{1+\zeta_1(\ell)\,\a_*}{1+\zeta_1(\ell)}\,,\qquad \zeta_1=\left(\frac{\ell_*}{\ell}\right)^2\,.
\ee
For both small and large $\zeta_1$ the kinetic term in Eq.\ \Eq{useful} dominates over the potential term, so the profile \Eq{prof2} defines an effective fractional charge $\a_{\rm eff}\approx \a_1(\ell)$ throughout the dimensional flow. With $N$ coefficients $\a_n$, $\a_N=1$, the effective fractional charge reads
\be\label{profN}
{\a_{N-1}(\ell):=\frac{1+\sum_{n=1}^{N-1}\zeta_n(\ell)\,\a_n}{1+\sum_{n=1}^{N-1}\zeta_n(\ell)}\,,\quad \zeta_n=\left(\frac{\ell_n}{\ell-\ell_{n-1}}\right)^2.}
\ee
In fact, this is nothing but the average $\langle\a\rangle$ of the coefficients $\a_n$ with respect to the weights $\zeta_n$.

\begin{figure*}
\centering
\includegraphics[width=6.9cm]{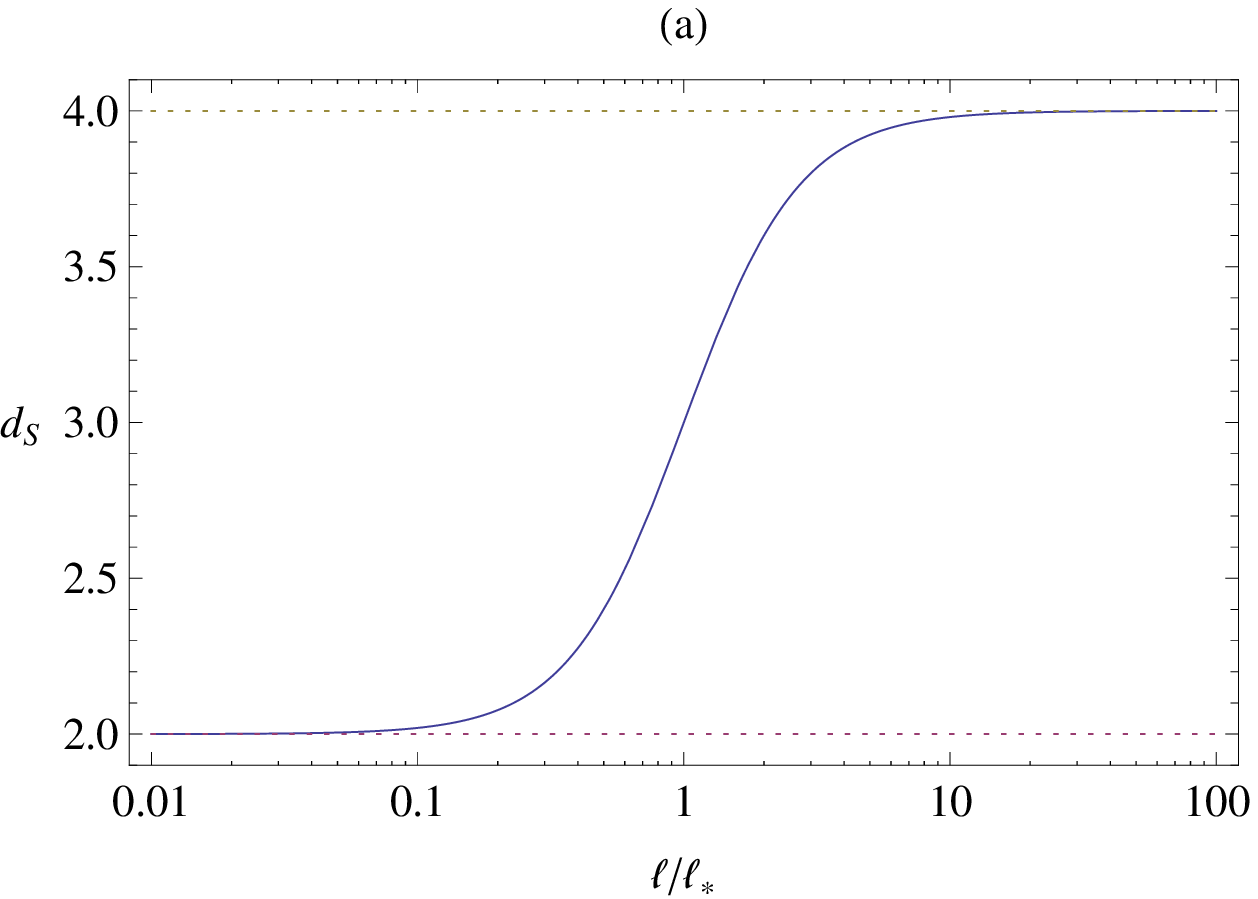}
\includegraphics[width=7cm]{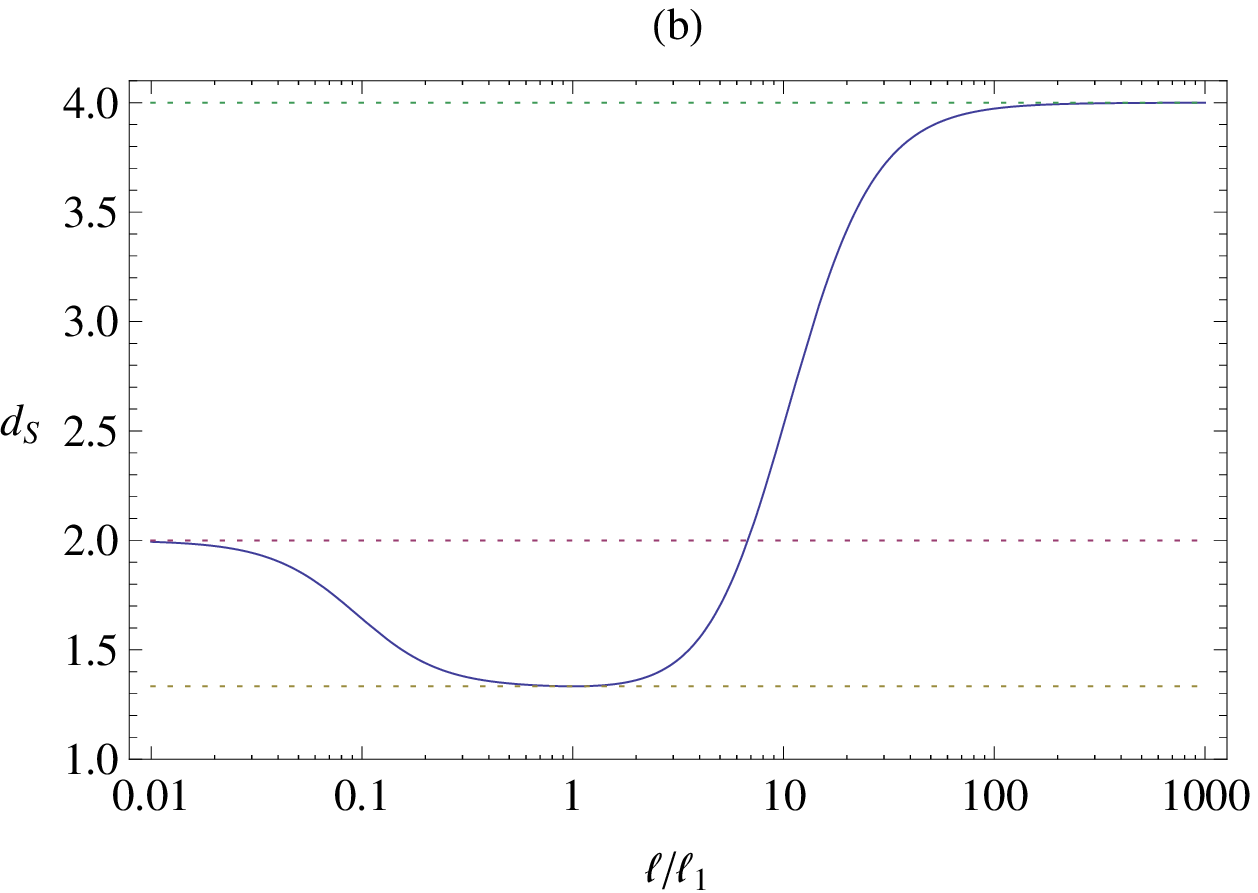}
\caption{\label{fig5} The spectral dimension $\ds(\ell)$ in $D=4$ for a multifractional model and normal diffusion ($\b=1=\g$, $\ds=\dh$) in four dimensions, with profile \Eq{prof1} (a) and \Eq{prof2} (b).}
\end{figure*}

The spectral dimension for $D=4$ in the one-scale case ($N=2$, $\a_*=1/2$) is shown in Fig.\ \ref{fig5}(a), in agreement with \ref{fig4}. A two-scale profile with $\a_1=1/2$, $\a_2=1/3$ and $\ell_2=10\ell_1$ is plotted in Fig.\ \ref{fig5}(b):
\bs\label{prof2}\ba
\a_2(\ell)&:=&\frac{1+\frac12\zeta_1(\ell)+\frac13\zeta_2(\ell)}{1+\zeta_1(\ell)+\zeta_2(\ell)}\,,\\
\zeta_1&=&\left(\frac{\ell_1}{\ell}\right)^2,\qquad \zeta_2=\left(\frac{\ell_2}{\ell-\ell_1}\right)^2.
\ea\es
At $\ell=0$ (beginning of the flow, UV critical point), $\ds=2$ in four dimensions. At $\ell\sim \ell_1$, the spectral dimension acquires the minimum value $\ds=4/3$. At scales $\ell\leq\ell_2$, the diffusion process corresponds to a \emph{recurrent} random walk \cite{bH}, where $\dw>\dh$ ($\ds<2$) and each site of the walk within a given radius is visited several times. Well above the larger critical scale, $\ell\gg\ell_2$, both dimensions hit the IR value $\sim 4$:
\be\label{iruv3}
\ds\sim\left\{ \begin{matrix} D\,,                & \ell\gg\ell_2\gg\ell_1\,\qquad \mbox{(IR)}\,,\\
                              D\a_2=\frac{D}{3}\,,& \qquad\ell_1\sim\ell\ll\ell_2\,\qquad \mbox{(intermediate)}\,,\\
                              D\a_1=\frac{D}{2}\,,& \ell\ll\ell_1\ll\ell_2\,\qquad \mbox{(UV)}\,.\end{matrix}\right. 
\ee

The two-scale spectral dimension of Fig.\ \ref{fig5}(b) reproduces the dimensional flow of QEG \cite{ReS11,fra6,CES}.


\section{Discussion}\label{disc}

Dimensional flow, the change of spacetime dimensionality with the scale, is a trademark of quantum-gravity scenarios which appears in different guises but bears a simple, limited set of general characteristics. We have listed and classified these characteristics in a portable fashion (i.e., independently of the quantum-gravity model) according to their degree of dependency on the details of the diffusion equation. While the number of asymptotic regimes (plateaux) for the spectral dimension, their values and positions are physical quantities related to the hierarchy of fundamental scales of the system, the intermediate, transient regimes connecting them depend on the specific realization of the diffusion equation. We have dissected the latter in its constituent elements and classified a number of ways to introduce a multiscale structure in it. While doing so, we pointed out how uncharted territory in quantum gravity can be explored using the maps of other branches of physics and mathematics such as diffusion and stochastic theory. Anomalous and multiscale diffusion equations, in fact, have been known since long ago in these disciplines. Some of the contributions of the present paper amount to link composite stochastic processes to multiscale spacetime geometries, as in the case of the fractional telegraph process (even in probability-theory literature, to the best of our knowledge, the latter seems not to be directly associated with a multiscale system), to ``locally'' characterize asymptotic regimes in the dimensional flow with specific stochastic processes, and to provide a qualitative (Sec.\ \ref{unun}) and quantitative (Sec.\ \ref{f6})  general analysis of dimensional flow. In parallel, we have improved the status of knowledge of multifractional spacetimes regarding fractional Laplacians, diffusion, and the analytic treatment of multiscale configurations.

The single-scale example is clear-cut in the way it illustrates how to associate a given dimensional flow or portions of it with specific types of stochastic processes. Suppose we have a quantum-gravity model with spacetime spectral dimension following the monotonic profile of Fig.\ \ref{fig1}. Different diffusion equations can give the same profile qualitatively, with the same asymptotics but a different slope in between [Figs.\ \ref{fig4} and \ref{fig5}(a) and, e.g., Fig.\ 1 of Ref.\ \cite{CaG}]. As we have seen, one such diffusion equation governs what is known as a {fractional telegraph process}, which is a telegraph process with Brownian time (Sec.\ \ref{tele}). In this stochastic process, the test particle experiences abrupt changes of speed at times governed by a Poisson law, in turn parametrized by a Wiener process (i.e., Brownian motion). At large scales, towards the upper right plateau of the figure, this process reduces to an ordinary Brownian motion. At small scales, in the lower left plateau, the process reduces to another composite process, called {iterated Brownian motion}, which is a Brownian motion with Brownian time. Iterated Brownian motion describes the diffusion of a Brownian particle in a fractal medium. Now, suppose the quantum-gravity model at hand does \emph{not} realize this particular diffusion equation (for instance, multifractional theory). Since dimensional flow is essentially the same, we can still interpret the one-scale anomalous diffusion equation as describing the diffusion of a particle in an irregular fractal medium, although the process is not really a fractional telegraph process. Nevertheless, the existence of a specific stochastic process in the same ``equivalence class'' of flows (in the example, the class of monotonic single-scale flows) gives a sharper characterization of the physics underlying diffusion in an anomalous spacetime. This is also true asymptotically; in the example, the asymptotic class of flows leading to $\ds\sim D/2$ is that of the IBM.

Through the exact and approximate solution of the various diffusion equations, we have replaced spotwise knowledge of asymptotic values of the spectral dimension with a continuous analytic control over dimensional flow in a model-independent fashion. In particular, control over multifractional spacetimes has been ameliorated. This should open up further advancements in other aspects of the field as well as its application as an effective framework to other theories of quantum gravity.


\begin{acknowledgments}
The author thanks A.\ Eichhorn, P.\ Ho\v{r}ava, and F.\ Saueressig for many stimulating discussions.
\end{acknowledgments}



\begin{thebibliography}{99}

\bibitem{tH93}  G.\ 't Hooft, \tia{Dimensional reduction in quantum gravity} in \emph{Salamfestschrift}, edited by A.\ Ali, J.\ Ellis, and S.\ Randjbar-Daemi (World Scientific, Singapore, 1993) [\oarX{gr-qc/9310026}].
\bibitem{Car09} S.~Carlip, \tia{Spontaneous dimensional reduction in short-distance quantum gravity?} 
  \doin{10.1063/1.3284402}{AIP Conf.\ Proc.}{1196}{72}{2009} [\arX{0909.3329}].
\bibitem{Car10} S.~Carlip, \tia{The small scale structure of spacetime} in \emph{Foundations of Space and Time}, edited by G.\ Ellis, J.\ Murugan, and A.\ Weltman (Cambridge University Press, Cambridge, U.K., 2012) [\arX{1009.1136}].
\bibitem{AJL4}  J.~Ambj{\o}rn, J.~Jurkiewicz, and R.~Loll, \tia{Spectral dimension of the universe}
\doin{10.1103/PhysRevLett.95.171301}{Phys.\ Rev.\ Lett.}{95}{171301}{2005} [\oarX{hep-th/0505113}].
\bibitem{BeH}   D.~Benedetti and J.~Henson, \tia{Spectral geometry as a probe of quantum spacetime} \doin{10.1103/PhysRevD.80.124036}{Phys.\ Rev.\ D}{80}{124036}{2009} [\arX{0911.0401}].
\bibitem{DJW1}  B.~Durhuus, T.~Jonsson, and J.F.~Wheater, \tia{Random walks on combs} \doin{10.1088/0305-4470/39/5/002}{J.\ Phys.\ A}{39}{1009}{2006} [\oarX{hep-th/0509191}].
\bibitem{AGW}   M.R.~Atkin, G.~Giasemidis, and J.F.~Wheater, \tia{Continuum random combs and scale dependent spectral dimension} \doin{10.1088/1751-8113/44/26/265001}{J.\ Phys.\ A}{44}{265001}{2011} [\arX{1101.4174}].
\bibitem{GWZ1}  G.~Giasemidis, J.F.~Wheater, and S.~Zohren, \tia{Dynamical dimensional reduction in toy models of $4D$ causal quantum gravity} \doin{10.1103/PhysRevD.86.081503}{Phys.\ Rev.\ D}{86}{081503(R)}{2012}
[\arX{1202.2710}].
\bibitem{GWZ2}  G.~Giasemidis, J.F.~Wheater, and S.~Zohren,
  \tia{Multigraph models for causal quantum gravity and scale dependent spectral dimension} \doin{10.1088/1751-8113/45/35/355001}{J.\ Phys.\ A}{45}{355001}{2012} [\arX{1202.6322}].
\bibitem{LaR5}  O.~Lauscher and M.~Reuter, \tia{Fractal spacetime structure
in asymptotically safe gravity} \doij{10.1088/1126-6708/2005/10/050}{J.\ High Energy Phys.}
{10}{050}{2005} [\oarX{hep-th/0508202}].
\bibitem{ReS11} M.~Reuter and F.~Saueressig, \tia{Fractal space-times under the microscope: a renormalization group view on Monte Carlo data} \doij{10.1007/JHEP12(2011)012}{J.\ High Energy Phys.}{12}{012}{2011} [\arX{1110.5224}]. 
\bibitem{Mod08} L.~Modesto, \tia{Fractal structure of loop quantum gravity}
\doin{10.1088/0264-9381/26/24/242002}{Class.\ Quantum Grav.}{26}{242002}{2009} [\arX{0812.2214}].
\bibitem{CaM}  F.~Caravelli and L.~Modesto, \tia{Fractal dimension in 3d spin-foams} \arX{0905.2170}.
\bibitem{MPM}  E.~Magliaro, C.~Perini, and L.~Modesto, \tia{Fractal space-time from spin-foams} \arX{0911.0437}.
\bibitem{COT}   G.\ Calcagni, D.\ Oriti, and J.\ Th\"urigen, \tia{Laplacians on discrete and quantum geometries} \arX{1208.0354}; and (in progress).
\bibitem{Hor3}  P.~Ho\v{r}ava, \tia{Spectral dimension of the universe in quantum gravity at a Lifshitz point} \doin{10.1103/PhysRevLett.102.161301}{Phys.\ Rev.\ Lett.}{102}{161301}{2009} [\arX{0902.3657}].
\bibitem{SVW1}  T.P.~Sotiriou, M.~Visser, and S.~Weinfurtner, \tia{Spectral dimension as a probe of the ultraviolet continuum regime of causal dynamical triangulations} \doin{10.1103/PhysRevLett.107.131303}{Phys.\ Rev.\ Lett.}{107}{131303}{2011} [\arX{1105.5646}].
\bibitem{Con06} A.\ Connes, \tia{Noncommutative geometry and the standard model with neutrino mixing}
 \doij{10.1088/1126-6708/2006/11/081}{J.\ High Energy Phys.}{11}{081}{2006} [\oarX{hep-th/0608226}].
\bibitem{CCM}   A.H.\ Chamseddine, A.\ Connes, and M.\ Marcolli, \ndoin{http://www.intlpress.com/ATMP/ATMP-issue_11_6.php}{Adv.\ Theor.\ Math.\ Phys.}{11}{991}{2007} [\oarX{hep-th/0610241}].
\bibitem{Ben08} D.\ Benedetti, \tia{Fractal properties of quantum spacetime} \doin{10.1103/PhysRevLett.102.111303}{Phys.\ Rev.\ Lett.}{102}{111303}{2009} [\arX{0811.1396}].
\bibitem{ACOS}  M.\ Arzano, G.\ Calcagni, D.\ Oriti, and M.\ Scalisi, \tia{Fractional
and noncommutative spacetimes} \doin{10.1103/PhysRevD.84.125002}{Phys.\ Rev.\ D}{84}{125002}{2011} [\arX{1107.5308}].
\bibitem{AA}    E.\ Alesci and M.\ Arzano, \tia{Anomalous dimension in semiclassical gravity} \doin{10.1016/j.physletb.2011.12.026}{Phys.\ Lett.\ B}{707}{272}{2012} [\arX{1108.1507}].
\bibitem{fra1}  G.\ Calcagni, \tia{Fractal universe and quantum gravity}
\doin{10.1103/PhysRevLett.104.251301}{Phys.\ Rev.\ Lett.}{104}{251301}{2010} [\arX{0912.3142}].
\bibitem{fra2}  G.\ Calcagni, \tia{Quantum field theory, gravity and cosmology in
a fractal universe} \doij{10.1007/JHEP03(2010)120}{J.\ High Energy Phys.}{03}{120}{2010} [\arX{1001.0571}].
\bibitem{fra3}  G.\ Calcagni, \tia{Gravity on a multifractal}
\doin{10.1016/j.physletb.2011.01.063}{Phys.\ Lett.\ B}{697}{251}{2011}
[\arX{1012.1244}]. 
\bibitem{fra4}  G.\ Calcagni, \tia{Discrete to continuum transition in
multifractal spacetimes}
 \doin{10.1103/PhysRevD.84.061501}{Phys.\ Rev.\ D}{84}{061501(R)}{2011}
[\arX{1106.0295}].
\bibitem{frc1}  G.\ Calcagni, \tia{Geometry of fractional spaces} {Adv.\ Theor.\ Math.\ Phys.} {\bf 16} (to be published) [\arX{1106.5787}].
\bibitem{frc2}  G.\ Calcagni, \tia{Geometry and field theory in multi-fractional
spacetime}
 \doij{10.1007/JHEP01(2012)065}{J.\ High Energy Phys.}{01}{065}{2012} [\arX{1107.5041}].
\bibitem{frc3}  G.\ Calcagni and G.\ Nardelli, \tia{Momentum transforms and Laplacians in fractional spaces} \arX{1202.5383}.
\bibitem{fra6}  G.\ Calcagni, \tia{Diffusion in quantum geometry} \doin{10.1103/PhysRevD.86.044021}{Phys.\ Rev.\ D}{86}{044021}{2012} [\arX{1204.2550}].
\bibitem{frc4}  G.\ Calcagni, G.\ Nardelli, and M.\ Scalisi, \tia{Quantum mechanics in fractional and other anomalous spacetimes} \doin{10.1063/1.4757647}{J.\ Math.\ Phys.}{53}{102110}{2012} [\arX{1207.4473}].
\bibitem{Mod11} L.\ Modesto, \tia{Super-renormalizable quantum gravity} \doin{10.1103/PhysRevD.86.044005}{Phys.\ Rev.\ D}{86}{044005}{2012} [\arX{1107.2403}].
\bibitem{AMM}   S.\ Alexander, A.\ Marcian\`o, and L.\ Modesto, \tia{The hidden quantum groups symmetry of super-renormalizable gravity} \doin{10.1103/PhysRevD.85.124030}{Phys.\ Rev.\ D}{85}{124030}{2012} [\arX{1202.1824}].
\bibitem{Mod12} L.\ Modesto, \tia{Super-renormalizable multidimensional quantum gravity} \arX{1202.3151}.
\bibitem{fra7}  G.\ Calcagni, \tia{Multi-fractional spacetimes, asymptotic safety and Ho\v{r}ava-Lifshitz gravity} \arX{1209.4376}.
\bibitem{CES}   G.\ Calcagni, A.\ Eichhorn, and F.\ Saueressig (in progress).
\bibitem{Fic55} A.\ Fick, \tia{\"Uber Diffusion} Pogg.\ Ann.\ Phys.\ Chem.\ {\bf 170}, 59 (1855).
\bibitem{Ein05} A.\ Einstein, \tia{\"Uber die von der molekularkinetischen Theorie der W\"arme geforderte Bewegung von in ruhenden Fl\"ussigkeiten suspendierten Teilchen}
  \doin{10.1002/andp.19053220806}{Ann.\ Phys.\ (Leipzig)}{322}{549}{1905}.
\bibitem{Smo06} M.\ von Smoluchowski, \tia{Zur kinetischen Theorie der Brownschen Molekularbewegung und der Suspensionen} \doin{10.1002/andp.19063261405}{Ann.\ Phys.\ (Leipzig)}{326}{756}{1906}.
\bibitem{SVW2}  T.P.\ Sotiriou, M.\ Visser, and S.\ Weinfurtner,
  \tia{From dispersion relations to spectral dimension ---and back again}
  \doin{10.1103/PhysRevD.84.104018}{Phys.\ Rev.\ D}{84}{104018}{2011} [\arX{1105.6098}].
\bibitem{CaG}   S.~Carlip and D.~Grumiller, \tia{Lower bound on the spectral dimension near a black hole}
 \doin{10.1103/PhysRevD.84.084029}{Phys.\ Rev.\ D}{84}{084029}{2011} [\arX{1108.4686}].
\bibitem{MoN}   L.~Modesto and P.~Nicolini, \tia{Spectral dimension of a quantum universe} \doin{10.1103/PhysRevD.81.104040}{Phys.\ Rev.\ D}{81}{104040}{2010} [\arX{0912.0220}].
\bibitem{SSN}   E.~Spallucci, A.~Smailagic, and P.~Nicolini, \tia{Trace anomaly in quantum spacetime manifold} \doin{10.1103/PhysRevD.73.084004}{Phys.\ Rev.\ D}{73}{084004}{2006} [\oarX{hep-th/0604094}].
\bibitem{RSnax} M.\ Reuter and F.\ Saueressig, \tia{Asymptotic safety, fractals, and cosmology} in \emph{Quantum Gravity and Quantum Cosmology}, edited by G.\ Calcagni, L.\ Papantonopoulos, G.\ Siopsis, and N.C.\ Tsamis  (Springer-Verlag, Berlin, 2013), \doin{10.1007/978-3-642-33036-0_8}{Lect.\ Notes Phys.}{863}{185}{2013} [\arX{1205.5431}].

\bibitem{KST}   A.A.~Kilbas, H.M.~Srivastava, and J.J.~Trujillo, \emph{Theory and Applications of Fractional Differential Equations} (Elsevier, Amsterdam, 2006).
\bibitem{Smi04} A.V.~Smilga, \tia{Benign versus malicious ghosts in higher-derivative theories}
  \doin{10.1016/j.nuclphysb.2004.10.037}{Nucl.\ Phys.\ B}{706}{598}{2005} [\oarX{hep-th/0407231}].
\bibitem{BMS}   T.~Biswas, A.~Mazumdar, and W.~Siegel,
  \tia{Bouncing universes in string-inspired gravity}
  \doij{10.1088/1475-7516/2006/03/009}{J.\ Cosmol.\ Astropart.\ Phys.}{03}{009}{2006} [\oarX{hep-th/0508194}].
\bibitem{BaK1}  N.~Barnaby and N.~Kamran,
  \tia{Dynamics with infinitely many derivatives: the Initial value problem}
  \doij{10.1088/1126-6708/2008/02/008}{J.\ High Energy Phys.}{02}{008}{2008} [\arX{0709.3968}].

\bibitem{ARAG}  J.M.\ Angulo, M.D.\ Ruiz-Medina, V.V.\ Anh, and W.\ Grecksch, \tia{Fractional diffusion and fractional heat equation}
  \doin{10.1239/aap/1013540349}{Adv.\ Appl.\ Probab.}{32}{1077}{2000}.
\bibitem{AAMR}  J.M.\ Angulo, V.V.\ Anh, R.\ McVinish, and M.D.\ Ruiz-Medina, \tia{Fractional kinetic equations driven by Gaussian or infinitely divisible noise}
  \doin{10.1239/aap/1118858630}{Adv.\ Appl.\ Probab.}{37}{366}{2005}.
\bibitem{BGG}   C.G.~Bollini, J.J.~Giambiagi, and A.~Gonz\'ales Dom\'{i}nguez, \tia{Analytic regularization and the divergences of quantum field theories} \doin{10.1007/BF02733756}{Nuovo Cimento}{31}{550}{1964}.
\bibitem{Gia91} J.J.~Giambiagi, \tia{Huygens' principle in $(2n+1)$-dimensions for nonlocal pseudodifferential operators of the type $\B^\alpha$} \doin{10.1007/BF02812498}{Nuovo Cim.\ A}{104}{1841}{1991}.
\bibitem{BGO}   C.G.~Bollini, J.J.~Giambiagi, and O.~Obregon, \tia{Criteria to fix the dimensionality corresponding to some higher derivative Lagrangians}
 \doin{10.1142/S0217732392000562}{Mod.\ Phys.\ Lett.\ A}{7}{593}{1992}.
\bibitem{doA92} R.L.P.~do Amaral and E.C.~Marino, \tia{Canonical quantization of theories containing fractional powers of the d'Alembertian operator} \doin{10.1088/0305-4470/25/19/026}{J.\ Phys.\ A}{25}{5183}{1992}.
\bibitem{BG}    C.G.~Bollini and J.J.~Giambiagi, \tia{Arbitrary powers of d'Alembertians and the Huygens' principle}
 \doin{10.1063/1.530263}{J.\ Math.\ Phys.}{34}{610}{1993}.
\bibitem{BOR}   D.G.~Barci, L.E.~Oxman, and M.~Rocca, \tia{Canonical quantization of non-local field equations}
 \doin{10.1142/S0217751X96001061}{Int.\ J.\ Mod.\ Phys.\ A}{11}{2111}{1996} [\oarX{hep-th/9503101}].
\bibitem{BBOR}  D.G.~Barci, C.G.~Bollini, L.E.~Oxman, and M.C.~Rocca, \tia{Lorentz-invariant pseudo-differential wave equations} \doin{10.1023/A:1026696132216}{Int.\ J.\ Theor.\ Phys.}{37}{3015}{1998}.

\bibitem{MeK}   R.~Metzler and J.~Klafter, \tia{The random walk's guide to anomalous diffusion: a fractional dynamics approach} \doin{10.1016/S0370-1573(00)00070-3}{Phys.\ Rep.}{339}{1}{2000}.
\bibitem{Zas3}  G.M.~Zaslavsky, \tia{Chaos, fractional kinetics, and anomalous transport}
 \doin{10.1016/S0370-1573(02)00331-9}{Phys.\ Rep.}{371}{461}{2002}.
\bibitem{MeK2}  E.\ Metzler and J.\ Klafter, \tia{The restaurant at the end of the random walk: recent developments in the description of anomalous transport by fractional dynamics}
  \doin{10.1088/0305-4470/37/31/R01}{J.\ Phys.\ A}{37}{R161}{2004}.
\bibitem{MCK}   R.\ Metzler, A.V.\ Chechkin, and J.\ Klafter, \tia{L\'evy statistics and anomalous transport: L\'evy flights and subdiffusion} in
\href{http://www.springerlink.com/content/978-0-387-75888-6}{\cob \it Encyclopedia of Complexity and Systems Science}, edited by R.A.\ Meyers (Springer, Amsterdam, 2009) [\arX{0706.3553}].
\bibitem{Sok12} I.M.\ Sokolov, \tia{Models of anomalous diffusion in crowded environments}
  \doin{10.1039/C2SM25701G}{Soft Matter}{8}{9043}{2012}.
\bibitem{MLP}   F.\ Mainardi, Y.\ Luchko, and G.\ Pagnini, \tia{The fundamental solution of the space-time fractional diffusion equation}  \ndoin{http://www.math.bas.bg/~fcaa}{Fractional Calculus Appl.\ Anal.}{4}{153}{2001} [\oarX{cond-mat/0702419}].
\bibitem{Kol40} A.N.\ Kolmogorov, Dokl.\ Akad.\ Nauk SSSR {\bf 26}, 115 (1940).
\bibitem{MaV}   B.B.\ Mandelbrot and J.W.\ Van Ness, \tia{Fractional Brownian motions, fractional noises and applications} \doin{10.1137/1010093}{SIAM Rev.}{10}{422}{1968}.

\bibitem{MBB1}  M.M.\ Meerschaert, D.A.\ Benson, and B.\ Baeumer, \tia{Multidimensional advection and fractional dispersion}
  \doin{10.1103/PhysRevE.59.5026}{Phys.\ Rev.\ E}{59}{5026}{1999}.
\bibitem{MBB2}  M.M.\ Meerschaert, D.A.\ Benson, and B.\ Baeumer, \tia{Operator L\'evy motion and multiscaling anomalous diffusion}
  \doin{10.1103/PhysRevE.63.021112}{Phys.\ Rev.\ E}{63}{21112}{2001}.
\bibitem{WeZ1}  H.\ Weitzner and G.M.\ Zaslavsky, \tia{Directional fractional kinetics}
  \doin{10.1063/1.1372514}{Chaos}{11}{384}{2001}.
\bibitem{SBMB}  R.\ Schumer, D.A.\ Benson, M.M.\ Meerschaert, and B.\ Baeumer, \tia{Multiscaling fractional advection-dispersion equations and their solutions}
  \doin{10.1029/2001WR001229}{Water Resour.\ Res.}{39}{1022}{2003}.
\bibitem{ZBMLS} Y.\ Zhang, D.A.\ Benson, M.M.\ Meerschaert, E.M.\ LaBolle, and H.-P.\ Scheffler, \tia{Random walk approximation of fractional-order multiscaling anomalous diffusion}
 \doin{10.1103/PhysRevE.74.026706}{Phys.\ Rev.\ E}{74}{026706}{2006}.
\bibitem{QiH} L.\ Yan-Qin and M.\ Jun-Hai, \tia{Exact solutions of a generalized multi-fractional nonlinear diffusion equation in radical symmetry}
  \doin{10.1088/0253-6102/52/5/20}{Commun.\ Theor.\ Phys.}{52}{857}{2009}.
\bibitem{RaG}   H.\ Rajaram and L.W.\ Gelhar, \tia{Plume-scale dependent dispersion in aquifers with a wide range of scales of heterogeneity}
  \doin{10.1029/95WR01723}{Water Resour.\ Res.}{31}{2469}{1995}.
\bibitem{ZhW}   H.\ Zhan and S.W.\ Wheatcraft, \tia{Macrodispersivity tensor for nonreactive solute transport in isotropic and anisotropic fractal porous media: analytical solutions}
  \doin{10.1029/95WR02282}{Water Resour.\ Res.}{32}{3461}{1996}.
\bibitem{HeP}   H.G.E.\ Hentschel and I.\ Procaccia. \tia{Relative diffusion in turbulent media: the fractal dimension of clouds}
  \doin{10.1103/PhysRevA.29.1461}{Phys.\ Rev.\ A}{29}{1461}{1984}.
\bibitem{OSP}   B.~O'Shaughnessy and I.~Procaccia, \tia{Analytical solutions for diffusion on fractal objects}
 \doin{10.1103/PhysRevLett.54.455}{ Phys.\ Rev.\ Lett.}{54}{455}{1985}.
\bibitem{MGN}   R.~Metzler, W.G.~Gl\"ockle, and T.F.~Nonnenmacher, \tia{Fractional model equation for anomalous diffusion} \doin{10.1016/0378-4371(94)90064-7}{Physica A}{211}{13}{1994}.
\bibitem{GR}    I.S.~Gradshteyn and I.M.~Ryzhik, {\it Table of Integrals, Series, and Products} (Academic Press, London, 2007).
\bibitem{SZ}   A.I.\ Saichev and G.M.\ Zaslavsky, \tia{Fractional kinetic equations: solutions and applications}
  \doin{10.1063/1.166272}{Chaos}{7}{753}{1997}.
\bibitem{WeZ2}  H.\ Weitzner and G.M.\ Zaslavsky, \tia{Some applications of fractional equations}
  \doin{10.1016/S1007-5704(03)00049-2}{Commun.\ Nonlinear Sci.\ Numer.\ Simul.}{8}{273}{2003} [\oarX{nlin/0212024}].
\bibitem{Nig86} R.R.\ Nigmatullin, \tia{The realization of the generalized transfer equation in a medium with fractal geometry}
  \doin{10.1002/pssb.2221330150}{Phys.\ Status Solidi B}{133}{425}{1986}.
\bibitem{Wis86} W.\ Wyss, \tia{The fractional diffusion equation}
  \doin{10.1063/1.527251}{J.\ Math.\ Phys.}{27}{2782}{1986}.
\bibitem{ScW}   W.R.\ Schneider and W.\ Wyss, \tia{Fractional diffusion and wave equations}
  \doin{10.1063/1.528578}{J.\ Math.\ Phys.}{30}{134}{1989}.
\bibitem{Ma95a} F.\ Mainardi, \tia{The time fractional diffusion-wave equation} {Izv.\ Vyssh.\ Uchebn.\ Zaved., Radiofiz.} {\bf 38}, 20 (1995) [\doin{10.1007/BF01051854}{Radiophys.\ Quantum Electron.}{38}{13}{1995}].
\bibitem{Mai96} F.\ Mainardi, \tia{The fundamental solutions for the fractional diffusion-wave equation}
  \doin{10.1016/0893-9659(96)00089-4}{Appl.\ Math.\ Lett.}{9}{23}{1996}.

\bibitem{BO1}   L.\ Beghin and E.\ Orsingher, \tia{The telegraph process stopped at stable-distributed times and its connection with the fractional telegraph equation} {Fractional Calculus Appl.\ Anal.} {\bf 6}, 187 (2003).
\bibitem{OB2}   E.\ Orsingher and L.\ Beghin, \tia{Fractional diffusion equations and processes with randomly varying time}
  \doin{10.1214/08-AOP401}{Ann.\ Probab.}{37}{206}{2009} [\arX{1102.4729}].

\bibitem{Fun79} T.\ Funaki, \tia{Probabilistic construction of the solution of some higher order parabolic differential equation}
  \ndoin{http://projecteuclid.org/euclid.pja/1195517312}{Proc.\ Jpn.\ Acad., Ser.\ A}{55}{176}{1979}.
\bibitem{DeM}   P.\ Deheuvels and D.M.\ Mason, \tia{A functional LIL approach to pointwise Bahadur-Kiefer theorems} in {\it Probability in Banach Spaces 8}, edited by R.M.\ Dudley, M.G.\ Hahn, and J.\ Kuelbs (Birkh\"auser, Boston, 1992). 
\bibitem{Bur92} K.\ Burdzy, \tia{Some path properties of iterated Brownian motion} in {\it Seminar on Stochastic Processes, 1992}, edited by E.\ \c{C}inlar, K.L.\ Chung, and M.J.\ Sharpe (Birkh\"auser, Boston, 1993). 
\bibitem{Bur94} K.\ Burdzy, \tia{Variation of iterated Brownian motion} in {\it Measure-Valued Processes, Stochastic Partial Differential Equations, and Interacting Systems}, edited by D.A.\ Dawson (AMS, Providence, 1994). 
\bibitem{KL1}   D.\ Khoshnevisan and T.M.\ Lewis, \tia{The uniform modulus of continuity of iterated Brownian motion}
  \doin{10.1007/BF02214652}{J.\ Theor.\ Probab.}{9}{317}{1996}.
\bibitem{KL2}   D.\ Khoshnevisan and T.M.\ Lewis, \tia{Chung's law of the iterated logarithm for iterated brownian motion} \ndoin{http://www.numdam.org/item?id=AIHPB_1996__32_3_349_0}{Ann.\ I.H.P.: Phys.\ Theor.}{32}{349}{1996}.
\bibitem{CCFR}  E.\ Cs\'aki, M.\ Cs\"org\H{o}, A.\ F\"oldes, and P.\ R\'ev\'esz, \tia{The local time of iterated Brownian motion}
  \doin{10.1007/BF02214084}{J.\ Theor.\ Probab.}{9}{717}{1996}.
\bibitem{KL3}   D.\ Khoshnevisan and T.M.\ Lewis, \tia{Stochastic calculus for Brownian motion on a Brownian fracture}
  \doin{10.1214/aoap/1029962807}{Ann.\ Appl.\ Probab.}{9}{629}{1999}.  
\bibitem{AZ}    H.\ Allouba and W.\ Zheng, \tia{Brownian-time processes: the PDE connection and the half-derivative generator}
  \doin{10.1214/aop/1015345772}{Ann.\ Probab.}{29}{1780}{2001} [\arX{1005.3801}].
\bibitem{All02} H.\ Allouba, \tia{ Brownian-time processes: the PDE connection II and the corresponding Feynman--Kac formula} 
  \doin{10.1090/S0002-9947-02-03074-X}{Trans.\ Am.\ Math.\ Soc.}{354}{4627}{2002}.
\bibitem{DeB04} R.D.\ DeBlassie, \tia{Iterated Brownian motion in an open set}
  \doin{10.1214/105051604000000404}{Ann.\ Appl.\ Probab.}{14}{1529}{2004}.
\bibitem{OB1}   E.\ Orsingher and L.\ Beghin, \tia{Time-fractional telegraph equations and telegraph processes with Brownian time}
  \doin{10.1007/s00440-003-0309-8}{Probab.\ Theory Relat.\ Fields}{128}{141}{2004}.
\bibitem{Nan05} E.\ Nane, \tia{Iterated Brownian motion in bounded domains in $\mathbb{R}^n$}
  \doin{10.1016/j.spa.2005.10.007}{Stochastic Proc.\ Appl.}{116}{905}{2006} [\oarX{math/0505026}].
\bibitem{BMN1}  B.\ Baeumer, M.M.\ Meerschaert, and E.\ Nane, \tia{Brownian subordinators and fractional Cauchy problems}
  \doin{10.1090/S0002-9947-09-04678-9}{Trans.\ Am.\ Math.\ Soc.}{361}{3915}{2009} [\arX{0705.0168}].
\bibitem{BO3}   L.\ Beghin and E.\ Orsingher, \tia{Iterated elastic Brownian motions and fractional diffusion equations}
  \doin{10.1016/j.spa.2008.10.001}{Stochastic Proc.\ Appl.}{119}{1975}{2009}.
\bibitem{Nan08} E.\ Nane, \tia{Stochastic solutions of a class of higher order Cauchy problems in $\mathbb{R}^d$}
  \doin{10.1142/S021949371000298X}{Stochastics Dyn.}{10}{341}{2010} [\arX{0809.4824}].
\bibitem{BMN2}  B.\ Baeumer, M.M.\ Meerschaert, and E.\ Nane, \tia{Space-time duality for fractional diffusion}
  \doin{10.1239/jap/1261670691}{J.\ Appl.\ Probab.}{46}{1100}{2009} [\arX{0904.1176}].  
\bibitem{BOS}   L.\ Beghin, E.\ Orsingher, and L.\ Sakhno, \tia{Equations of mathematical physics and compositions of Brownian and Cauchy processes}
  \doin{10.1080/07362994.2011.581071}{Stoch.\ Anal.\ Appl.}{29}{551}{2011} [\arX{1008.0928}].

\bibitem{OZ}    E.\ Orsingher and X.\ Zhao, \tia{Iterated processes and their applications to higher-order differential equations}
  \doin{10.1007/BF02650660}{Acta Math.\ Sinica}{15}{173}{1999}.
\bibitem{HO}    K.J.\ Hochberg and E.\ Orsingher, \tia{Composition of stochastic processes governed by higher-order parabolic and hyperbolic equations}
  \doin{10.1007/BF02214661}{J.\ Theor.\ Probab.}{9}{511}{1996}.

%
\bibitem{BuK}   K.\ Burdzy and D.\ Khoshnevisan, \tia{Brownian motion in a Brownian crack}
  \doin{10.1214/aoap/1028903448}{Ann.\ Appl.\ Probab.}{8}{708}{1998}.
\bibitem{GiR}   M.\ Giona and H.E.\ Roman, \tia{Fractional diffusion equation for transport phenomena in random media}
\doin{10.1016/0378-4371(92)90441-R}{Physica A}{185}{87}{1992}.
\bibitem{MeN}   R.\ Metzler and T.F.\ Nonnenmacher, \tia{Fractional diffusion: exact representations of spectral functions} \doin{10.1088/0305-4470/30/4/011}{J.\ Phys.\ A}{30}{1089}{1997}.
  
\bibitem{Kri60} V.Yu.\ Krylov, \tia{Some properties of the distribution corresponding to the equation $\p u/\p t = (-1)^{n+1} \p^{2q}u/\p x^{2q}$} {Sov.\ Math.\ Dokl.} {\bf 1}, 260 (1960).
\bibitem{Hoc79} K.J.\ Hochberg, \tia{A signed measure on path space related to Wiener measure}
  \doin{10.1214/aop/1176995529}{Ann.\ Probab.}{6}{433}{1978}.
\bibitem{BO2}   L.\ Beghin and E.\ Orsingher, \tia{The distribution of the local time for ``pseudoprocesses'' and its connection with fractional diffusion equations}
  \doin{10.1016/j.spa.2005.02.001}{Stochastic Proc.\ Appl.}{115}{1017}{2005}.

\bibitem{GrP}   P.\ Grassberger and I.\ Procaccia, \tia{Dimensions and entropies of strange attractors from a fluctuating dynamics approach}
 \doin{10.1016/0167-2789(84)90269-0}{Physica D}{13}{34}{1984}.
\bibitem{JKLOS} M.H.\ Jensen, L.P.\ Kadanoff, A.\ Libchaber, I.\ Procaccia, and J.\ Stavans, \tia{Global universality at the onset of chaos: results of a forced Rayleigh--B\'enard experiment}
  \doin{10.1103/PhysRevLett.55.2798}{Phys.\ Rev.\ Lett.}{55}{2798}{1985}.
\bibitem{HJKPS} T.C.\ Halsey, M.H.\ Jensen, L.P.\ Kadanoff, I.\ Procaccia, and B.I.\ Shraiman, \tia{Fractal measures and their singularities: the characterization of strange sets}
  \doin{10.1103/PhysRevA.33.1141}{Phys.\ Rev.\ A}{33}{1141}{1986};
  \doin{10.1103/PhysRevA.34.1601}{}{34}{1601}{1986}.
\bibitem{PV}    G.\ Paladin and A.\ Vulpiani, \tia{Anomalous scaling laws in multifractal objects}
  \doin{10.1016/0370-1573(87)90110-4}{Phys.\ Rep.}{156}{147}{1987}.
\bibitem{Har01} D.~Harte, {\it Multifractals: Theory and Applications} (Chapman \& Hall/CRC, Boca Raton, FL, 2001).
\bibitem{Fal03} K.~Falconer, \textit{Fractal Geometry} (Wiley, New York, 2003).
\bibitem{Zas00} G.M.\ Zaslavsky, \tia{Multifractional kinetics}
  \doin{10.1016/S0378-4371(00)00441-6}{Physica A}{288}{431}{2000}.
\bibitem{Zas1}  G.M.\ Zaslavsky, \tia{Renormalization group theory of anomalous transport in systems with Hamiltonian chaos} \doin{10.1063/1.166054}{Chaos}{4}{25}{1994}.

\bibitem{Cap69} M.\ Caputo, \emph{Elasticit\`a e Dissipazione} (Zanichelli, Bologna, 1969) (in Italian).
\bibitem{Cap95} M.\ Caputo, \tia{Mean fractional-order-derivatives, differential equations and filters}
  \doin{10.1007/BF02826009}{Ann.\ Univ.\ Ferrara VII}{41}{73}{1995}.
\bibitem{BT1}   R.L.\ Bagley and P.J.\ Torvik, \tia{On the existence of the order domain and
the solution of distributed order equations: I} Int.\ J.\ Appl.\ Math.\ {\bf 2}, 865 (2000). 
\bibitem{BT2}   R.L.\ Bagley and P.J.\ Torvik, \tia{On the existence of the order domain and
the solution of distributed order equations: II} Int.\ J.\ Appl.\ Math.\ {\bf 2}, 965 (2000).
\bibitem{CGS}   A.V.\ Chechkin, R.\ Gorenflo, and I.M.\ Sokolov, \tia{Retarding subdiffusion and accelerating superdiffusion governed by distributed-order fractional diffusion equations} \doin{10.1103/PhysRevE.66.046129}{Phys.\ Rev.\ E}{66}{046129}{2002} [\oarX{cond-mat/0202213}].
\bibitem{LoH02} C.F.\ Lorenzo and T.T.\ Hartley, \tia{Variable order and distributed order fractional operators} \doin{}{Nonlinear Dyn.}{29}{57}{2002}.
\bibitem{Koc1}  A.N.\ Kochubei, \tia{Distributed order calculus and equations of ultraslow diffusion}
\doin{10.1016/j.jmaa.2007.08.024}{J.\ Math.\ Anal.\ Appl.}{340}{252}{2008} [\oarX{math-ph/0703046}].
\bibitem{Koc2}  A.N.\ Kochubei, \tia{Distributed-order calculus: an operator-theoretic interpretation}
  \doin{10.1007/s11253-008-0076-x}{Ukr.\ Math.\ J.}{60}{551}{2008} [\arX{0710.1710}].
\bibitem{Cap01} M.\ Caputo, \tia{Distributed order differential equations modelling dielectric
induction and diffusion}  \ndoin{www.math.bas.bg/~fcaa}{Fractional Calculus Appl.\ Anal.}{4}{421}{2001}.
\bibitem{CKS}   A.V.\ Chechkin, J.\ Klafter, and I.M.\ Sokolov, \tia{Fractional Fokker--Planck equation for ultraslow kinetics} \doin{10.1209/epl/i2003-00539-0}{Europhys.\ Lett.}{63}{326}{2003} [\oarX{cond-mat/0301487}].
\bibitem{Nab04} M.\ Naber, \tia{Distributed order fractional sub-diffusion} \doin{10.1142/S0218348X04002410}{Fractals}{12}{23}{2004} [\oarX{math-ph/0311047}].
\bibitem{MPG}   F.\ Mainardi, G.\ Pagnini, and R.\ Gorenflo, \tia{Some aspects of fractional diffusion equations of single and distributed order} \doin{10.1016/j.amc.2006.08.126}{Appl.\ Math.\ Comput.}{187}{295}{2007}	[\arX{0711.4261}].
\bibitem{CGGKS} A.V.\ Chechkin, V.Yu.\ Gonchar, R.\ Gorenflo, N.\ Korabel, and I.M.\ Sokolov, \tia{Generalized fractional diffusion equations for accelerating subdiffusion and truncated L\'evy flights} \doin{10.1103/PhysRevE.78.021111}{Phys.\ Rev.\ E}{78}{021111}{2008}.
\bibitem{CSK}   A.\ Chechkin, I.M.\ Sokolov, and J.\ Klafter, \tia{Natural and modified forms of distributed-order fractional diffusion equations} in \href{http://dx.doi.org/10.1142/9789814340595_0005}{\cob\it Fractional Dynamics: Recent Advances}, edited by J.\ Klafter, S.C.\ Lim, and R.\ Metzler (World Scientific, Singapore, 2011). 

\bibitem{Gol51} S.\ Goldstein, \tia{On diffusion by discontinuous movements and the telegraph equation}
  \doin{10.1093/qjmam/4.2.129}{Q.\ J.\ Mech.\ Appl.\ Math.}{4}{129}{1951}.
\bibitem{Kac74} M.\ Kac, \tia{A stochastic model related to the telegrapher's equation} 
  \doin{10.1216/RMJ-1974-4-3-497}{Rocky Mountain J.\ Math.}{4}{497}{1974}.
\bibitem{Ors85} E.\ Orsingher, \tia{Hyperbolic equations arising in random models}
  \doin{10.1016/0304-4149(85)90379-5}{Stochastic Proc.\ Appl.}{21}{93}{1985}.
\bibitem{Ors90} E.\ Orsingher, \tia{Probability law, flow function, maximum distribution of wavegoverned random motions and their connection with Kirchhoff's law}
  \doin{10.1016/0304-4149(90)90056-X}{Stochastic Proc.\ Appl.}{34}{49}{1990}.
\bibitem{SMH0}  R.K.\ Saxena, A.M.\ Mathai, and H.J.\ Haubold, \tia{Reaction-diffusion systems and non-linear waves}
  \doin{10.1007/s10509-006-9190-0}{Astrophys.\ Space Sci.}{305}{297}{2006} [\oarX{math/0604474}].
  
%
\bibitem{bH}    D.\ ben-Avraham and S.\ Havlin, \textit{Diffusion and Reactions in Fractals and Disordered Systems} (Cambridge University Press, Cambridge, UK, 2000).
\end{thebibliography}
\end{document}